%% file: lp95lnf.tex
\input phyzelv
\input defs

\input subsec

\input tables
\input epsfig

\let\cl=\centerline

 \def\plm{\ifm{\pm}}

\def\undertext#1{$\underline{\hbox{#1}}$}

\catcode`@=11 
\abovedisplayskip 2pt plus .1pt minus .2pt
\abovedisplayshortskip 2pt plus .1pt minus .1pt
\belowdisplayskip 2pt plus .1pt minus .1pt
\belowdisplayshortskip 2pt plus .1pt minus .1pt

\def\chapter#1{\par \penalty-300 \vskip12pt
   \spacecheck\chapterminspace
   \chapterreset
\noindent 
   {\bf\chapterlabel \ #1}\par
   \nobreak\vskip8pt \penalty 30000
   \wlog{\string\chapter\ \chapterlabel} }
\def\sectionlabel{\number\sectionnumber.}
\def\section#1{\par \ifnum\the\lastpenalty=30000\else
   \penalty-200\vskip\sectionskip \spacecheck\sectionminspace\fi
   \wlog{\string\section\ \chapterlabel \the\sectionnumber}
   \global\advance\sectionnumber by 1  \noindent
   {\it\chapterlabel \sectionlabel \ #1}\par
   \nobreak\vskip\headskip \penalty 30000 }

\catcode`@=12 

\hsize=15.2 cm \hoffset=0.5cm
\vsize=21.8 cm

\Frontpage

\vbox to 3.5cm{
\vglue1cm
\line{\hfill\undertext{LNF-95/054 (P)}}
\line{\hfill 16 October 1995}
\vfill}

\vglue 4mm
\vbox to5cm{
{\fourteenpoint
\centerline{\bf CP and RARE DECAYS}}
\vglue4mm
\baselineskip 15pt
\cl{PAOLO FRANZINI}
\vglue 4mm
\cl{\it Universit\`a di Roma I, La Sapienza}
\cl{\it Piazzale A. Moro 2, I-00185, Roma}
\cl{E-mail:Paolo@VAXLNF.LNF.INFN.IT}
}

\baselineskip 13pt
\vglue 6mm
\cl{\bf Abstract}
\parshape 1 12mm 128mm
\noindent
$CP$ violation experiments, new measurements of the parameters of the 
neutral \K\ system and searches for rare decays are summarized.
Perspectives
for the near future are presented.
\vglue2cm
\noindent

\vfill

{\it Invited Talk at the XVII International Symposium on Lepton-Photon 
Interactions, August 10-15, 1995, Beijing, China. To be Published in the 
Proceedings.}

\eject

\twelvepoint
\baselineskip=14pt
\def\C{$C$}
\def\P{$P$}

\def\noc{\rlap{$C$}\raise.3mm\hbox{\kern.5mm\\\kern.7mm}}
\def\nop{\rlap{$P$}\raise.3mm\hbox{\kern.5mm\\\kern.7mm}}
\def\noT{\rlap{$T$}\raise.3mm\hbox{\kern.5mm\\\kern.7mm}}

\def\nocp{\noc\nop}
\def\nocpt{\noc\nop\noT}
\def\B{\ifm{BR}}
\def\prl{Phys. Rev. Lett}

\vglue 5cm

\chapter{Introduction}

The study of \K\ mesons, begun more than 50 years ago,
has been central to the development of the standard model.
$CP$ violation was discovered in '64\rlap,\Ref\fitch{J. H. Christenson 
\etal, Phys. Rev. Lett. {\bf13} (1964) 138.}
through the observation of the unexpected decay \kl\to\pic. Since 
then, experiments searching for a 
difference in $\eta_{+-}$ and $\eta_{00}$ have been going on.
The complex amplitude ratios are defined in the standard notation 
as:\Ref\etaref{T. T. Wu and C. N. Yang \prl. {\bf 13} (1964) 380.}
$$\eqalign{
{A(\kl\to\pic)\over A(\ks\to\pic)}=&|\eta_{+-}|e^{i\phi_{+-}}=\eps+\eps'\cr
{A(\kl\to\pio)\over A(\ks\to\pio)}=&|\eta_{00}|e^{i\phi_{00}}=\eps-2\eps'\cr
}.$$

The classical measurable quantity ${\cal R}$, the so called double ratio
of the four rates for \kls\to\pio,\pic\ as defined below, is related to
\eps, $\eps'$ by 
$${\cal R}=\Big|{\eta_{00}\over\eta_{+-}}\Big|^2=1-6\rep.$$ 
Observation of \rep$\ne$0 is proof of ``direct'' $CP$ violation, \ie\
that the amplitude for $|\Delta S|$=1, $CP$ violating transitions
$A(K_2\to2\pi)\ne0$. 

All observations of \C\P\ violation, \nocp\ for short, \ie\ the decays
\kl\to2$\pi$, \pic\gam\ and the charge asymmetries in $K_{\ell3}$ decays are
examples of so called ``indirect'' violation, due to $|\Delta S|$=2
\ko$\leftrightarrow$\kob\ transitions introducing a small $CP$ impurity in the
mass eigenstates 
$$\ks\ab(\K_1+\eps\K_2)/\sqrt2,\quad 
\kl\ab (\K_2+\eps\K_1)/\sqrt2$$ 
where $\K_1$ and $\K_2$ 
are the $CP$ even and odd superposition of \ko, \kob\ and \eps\ab\pt2,-3,.

There is no new information on direct \nocp\ and we are still 
confronted with a slightly unsatisfactory experimental 
situation:\Ref\epseps{L. K. Gibbons \etal\ (E731), \prl. {\bf 70} (1993) 
1203; G. D. Barr \etal\ (NA31), Phys. Lett. {bf317} (1993) 233.}
$$\eqalign{
\rep=&\hbox{\pt(7.4$\pm$5.9),-4,}\cr
\rep=&\hbox{\pt(23$\pm$6.5),-4,}\cr
}$$
Taking the Particle Data Group's\Ref\pdg{Review of Particle Properties, 
Phys. Rev. {\bf D50} (1994) 1173.} (PDG94) average at face value, 
we could say that the confidence level that $0<\rep<\hbox{\pt3,-3,}$ is 94\%.
I will come back to the future prospects in this field.

A fundamental task of experimental physics today is the 
determination of the four parameters of the CKM mixing matrix, 
including the phase which results in \nocp. A knowledge of all 
parameters is required to confront experiments. Rather, many experiments 
are necessary to complete our knowledge of the parameters and prove 
the uniqueness of the model or maybe finally break beyond it.
As it happens rare \K\ decays can be crucial to this task.
I will therefore discuss the following topics:
new measurements of \ks, \kl\ parameters and searches for 
symmetry violations; new rare \K\ decay results;
other searches for \nocp\ and \noT. I will also briefly describe
perspectives for developments in the near future.

\chapter{New Measurements of the Neutral Kaon Properties}

\section{CPLEAR}
The CPLEAR experiment\Ref\CPlear{R. Adler \etal, CPLEAR Coll., to be 
submitted to NIM {\bf A}.} studies 
neutral \K\ mesons produced in equal numbers in proton-antiproton 
annihilations at rest:
$$\eqalign{p\bar p\to&K^-\pi^+\ko\quad\hbox{BR \pt2,-3,}\cr
                  \to&K^+\pi^-\kob\quad\hbox{BR \pt2,-3,}\cr}$$
\REFS\cpleara{R. Adler \etal, CPLEAR Coll., preprint CERN-PPE/95-103.}
\REFSCON\cplearb{R. Adler \etal, CPLEAR Coll., preprint 
CERN-PPE/95-107.}
The charge of $K^\pm(\pi^\pm)$ tags the strangeness $S$ of the neutral 
\K\ at $t$=0. They have recently presented several new results\refsend
from studying \pic, \pic\po\ and $\pi^\pm\ell^\mp\bar\nu(\nu)$ 
final states.
Their measurement of the \kl--\ks\ mass difference $\Delta m$
is independent of the value of $\phi_{+-}$, unlike in most other 
experiments. They have improved limits on the
possible violation of the $\Delta S=\Delta Q$ rule, quantified by the 
amplitude's ratio
$x=A(\Delta S=-\Delta Q)/A(\Delta S=\Delta Q)$, without assuming 
$CPT$ invariance. A direct test of $CPT$ invariance has also been 
obtained.
The data require small corrections for background asymmetry \ab1\%,  
differences in tagging efficiency, $\varepsilon(K^+\pi^-)-
\varepsilon(K^-\pi^+)$\ab10\up{-3} and in detection,
$\varepsilon(\pi^+e^-)-\varepsilon(\pi^-e^+)$\ab\pt3,-3,. They also 
correct for some regeneration in the detector.

\subsec{$\ko(\kob)\to e^+ (e^-)$}
Of particular interest are the study of the decays 
$\ko(\kob)\to e^+ (e^-)$. One can define the four decay intensities:
$$\eqalign{
I^+(t)\hbox{ for }&\ko\to e^+\cr
\overline I^-(t)\hbox{ for }&\kob\to e^-\cr}
\Big\}\eqalign{\Delta S=0\quad\cr}
\eqalign{
\overline I^+(t)\hbox{ for }&\kob\to e^+\cr
I^-(t)\hbox{ for }&\ko\to e^-\cr}
\Big\}\eqalign{&|\Delta S|=2\cr}$$
where $\Delta S=0,2$ mean that the strangeness of the decaying \K\ is 
the same as it was at $t$=0 or has changed from \ko-\kob\ mixing. One 
can then define four asymmetries:
$$\eqalign{
A_1(t)&={I^+(t)+\overline I^-(t)-(\overline I^+(t)+I^-(t)) \over 
I^+(t)+\overline I^-(t)+\overline I^+(t)+I^-(t)}\cr
A_2(t)&={\overline I^-(t)+\overline I^+(t)-(I^+(t)+I^-(t)) \over 
\overline I^-(t)+\overline I^+(t)+I^+(t)+I^-(t)}\cr}$$
$$A_T(t)={\overline I^+(t)-I^-(t)\over \overline I^+(t)+I^-(t)},\quad
a_{CPT}(t)={\overline I^-(t)-I^+(t)\over \overline I^-(t)+I^+(t)}$$
From the time dependence of $A_1$ they obtain:
$\Delta m=(0.5274\pm0.0029\pm0.0005)\x10^{10}$ s\up{-1}, a result which 
is independent of $\phi_{+-}$ and $\Re x=(12.4\pm11.9\pm6.9)\x10^{-3}$, 
without assuming $CPT$. From $A_2$ and assuming $CPT$ they obtain $\Im x
=(4.8\pm4.3)\x10^{-3}$, a result \ab5 times more stringent than the 
PDG94 
world average. $A_T$ gives a direct measurement of $T$ violation. 
Assuming $CPT$, the 
expected value for $A_T$ is \pt6.52,-3,. The CPLEAR result 
is $A_T=(6.3\pm2.1\pm1.8)\x10^{-3}$. From a study of the $CPT$ violating 
asymmetry, $A_{CPT}(t)$, they obtain
$\Re\delta_{CPT}=(0.07\pm0.53\pm0.45)\x10^{-3}$. We will come 
back later to the definition of $\delta_{CPT}$.
\FIG\kkdec
\FIG\decdif
\vbox{\cl{\epsfig{file=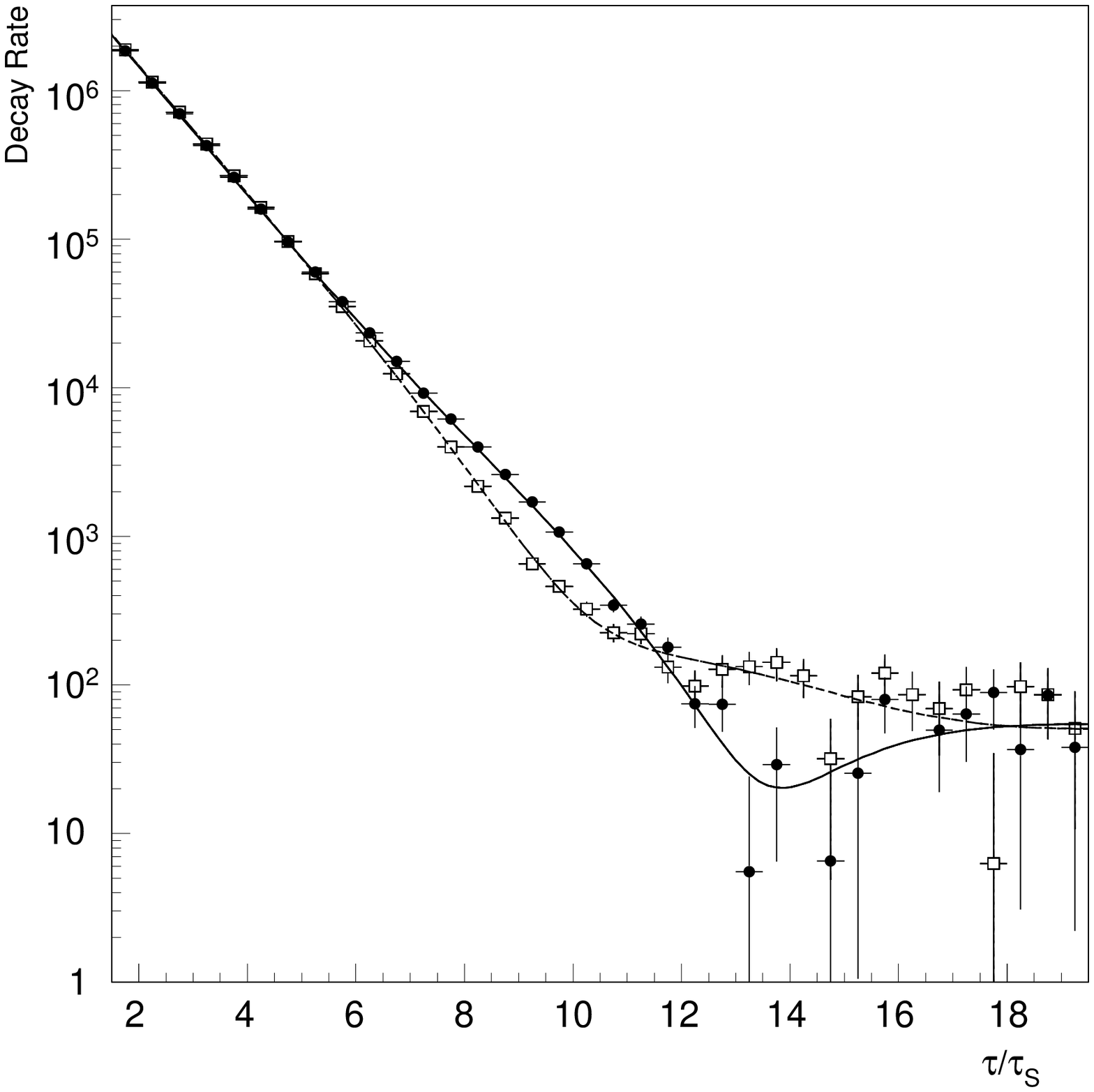,width=10cm}}
\cl{\tenpoint{\bf Fig. \chapterlabel\kkdec.} \ Decay distributions for 
\ko and \kob.}}
\vglue2mm
\subsec{\pic\po\ Final States}
Studies of \ko-\kob\to\pic\po\ decays give the results $\Re\eta_{+-0}=
(-4\pm17\pm3)\x10^{-3}$ and $\Im\eta_{+-0}=(-16\pm20\pm8)\x10^{-3}$, 
where $\eta_{+-0}=A(\kl\to\pic\po)/A(\ks\to\pic\po)$. 
By setting $\Re\eta_{+-0}=\Re\eta_{+-}$ they obtain $\Im(\eta_{+-0}=(-11
\pm14\pm8)\x10^{-3}$. These results are significantly more precise than any 
previous ones. 
\subsec{\pic\ Final State}
Finally from an analysis of \pt1.6,7, \pic\ decays of \ko\ 
and \kob\ they determine $|\eta_{+-}|=(2.312\pm0.043\pm0.03\pm0.011_{
\tau_S})\x10^{-3}$ and 
$\phi_{+-}=42.6\deg\pm0.9\deg\pm0.6\deg\pm0.9\deg_{\Delta m}$.
The {\it errors} in the values quoted reflect uncertainties in the 
knowledge of the \ks\ lifetime and the \ks--\kl\ mass difference, 
respectively. Fig. 
\chapterlabel\kkdec\ shows the decay intensities of \ko\ and \kob, while fig. 
\chapterlabel\decdif\ is a plot of the time dependent asymmetry $A_{+-}=
\big(I(\kob\to\pic)-\alpha I(\ko\to\pic)\big)/\big(I(\kob\to\pic)+
\alpha I(\ko\to\pic)\big)$. Most systematics cancel in the ratio and the 
residual difference in efficiencies for \ko\ and \kob\ decays is 
determined from a fit to the same data: $\alpha=0.9989\pm0.0006$.

\vbox{\cl{\epsfig{file=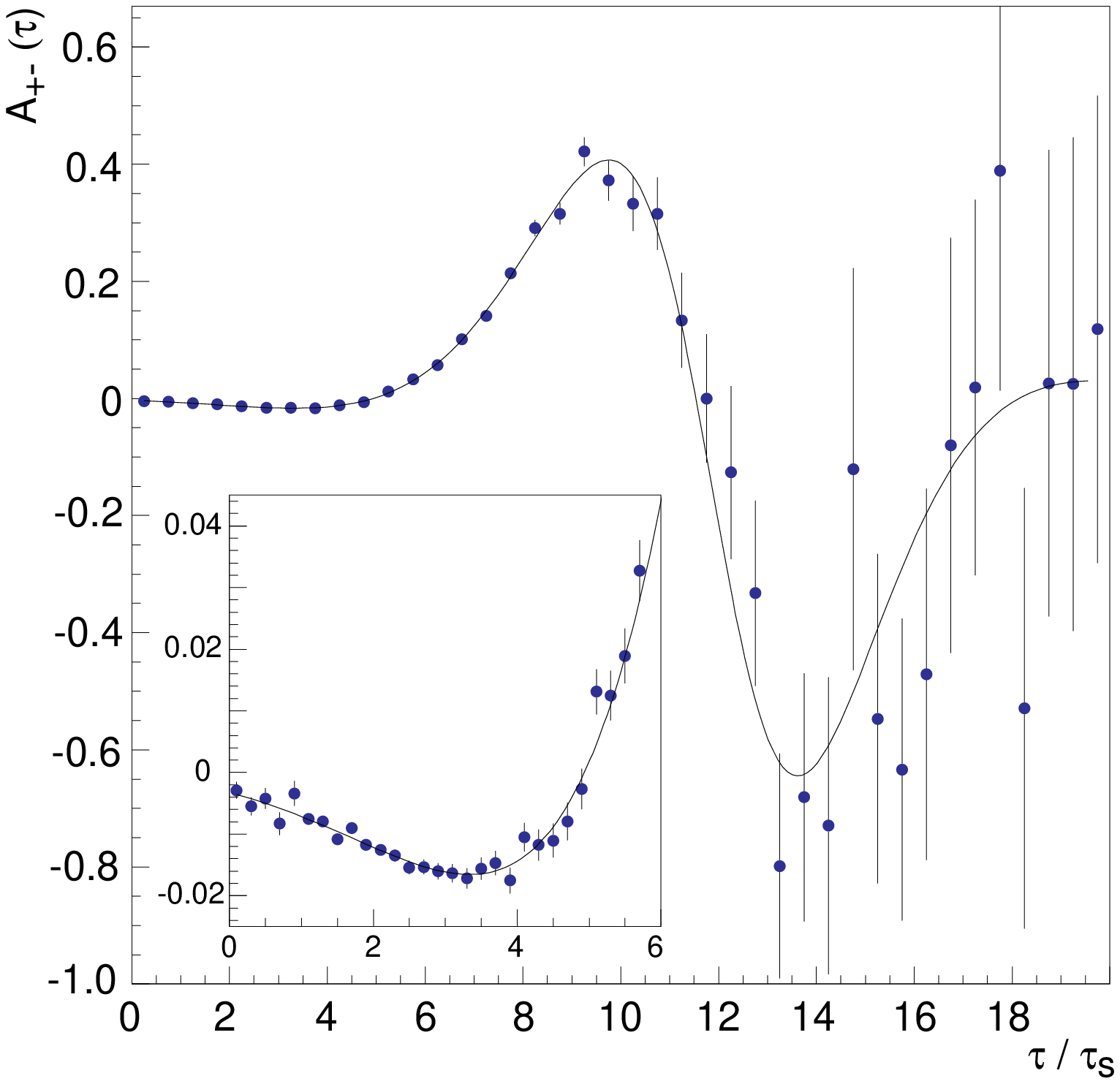,width=10cm}}
\cl{\tenpoint{\bf Fig. \chapterlabel\decdif.} \ Difference of decay 
distributions for \ko and \kob.}}

\section{E621 at FNAL}
For completeness the following new results for \K\to\pic\po\ must 
be mentioned\rlap.\Ref\pmzero{Y. Zou \etal, Phys. Let. {\bf B329} (1994) 
519; G. B. Thomson \etal, Phys. Lett. {\bf B337} (1994) 411.}
In this experiment the \C\P\ conserving amplitude 
$A$(\ks\to\pic\po) is measured, obtaining
$$\eqalign{|\rho_{\pic\po}|&=\Big|{A(\ks\to\pic\po,I=2)
\over A(\kl\to\pic\po)}\Big|=0.035^{+0.019}_{-0.011}\pm0.004\cr
\phi_\rho&=-59\deg\pm48\deg\cr
{\rm BR}(\ks\to\pic\po)&=(3.9^{+0.54+0.8}_{-1.8-0.7})\x 10^{-7}\cr
\Im(\eta_{+-0})&=-0.015\pm0.017\pm0.025,\quad
          {\rm assuming}\ \Re(\eta_{+-0})=\Re(\eps).\cr}$$

\section{E773 at FNAL}
E773 is essentially the old E731 setup, with minor improvements.
New results have been obtained on $\Delta m$, $\tau_S$, 
$\phi_{00}-\phi_{+-}$ and $\phi_{+-}$ from a study of \K\to\pic, \pio\ 
decays\rlap.\Ref\Schwin{B. Schwingenheuer \etal. \prl. {\bf74} (1995) 
4376.}
From a study of \pic\gam\ final states, $|\eta_{+-\gam}|$ and 
$\phi_{+-\gam}$ are obtained\rlap.\Ref\pmgam{J. N. Mathews \etal. \prl. 
{\bf75} (1995) 2803.}

\subsec{Two Pion Final States}
This study of \K\to$\pi\pi$ is a classic experiment where one beats
the amplitude $A(\kl\to\pi\pi\big]_i)$=$\eta_iA(\ks\to\pi\pi)$ with the 
coherently
regenerated \ks\to$\pi\pi$ amplitude $\rho A(\ks\to\pi\pi)$, resulting in
the decay intensity
$$\eqalign{I(t)=&|\rho|^2e^{-\Gamma_S t}+|\eta|^2e^{-\Gamma_L t}+\cr
&\qquad 2|\rho||\eta|e^{-\Gamma t}\cos(\Delta mt+\phi_\rho-\phi_{+-})\cr
}$$

Measurements of the time dependence of $I$ for the \pic\ final state
yields $\Gamma_S$, $\Gamma_L$, $\Delta m$ and $\phi_{+-}$. They give the 
following results:
$\tau_S=(0.8941\pm0.0014\pm0.009)$\x10\up{-10} s
setting $\phi_{+-}=\phi_{SW}=\tan^{-1}2\Delta m/\Delta\Gamma$ and
floating $\Delta m$;  
$\Delta m=(0.5297\pm0.0030\pm0.0022)$\x10\up{10} s\up{-1} 
using for $\tau_S$ the PDG94 value, leaving $\phi_{+-}$ free in the fit;
$\phi_{+-}=43.53\deg\pm0.58\deg\pm0.40\deg$, using for $\tau_S$ 
the PDG94 value and for the mass difference the combined values of E731 
and E773, $\Delta m=(0.5282\pm0.0030)\x10^{10}$ s\up{-1}. Including the 
uncertainties on $\Delta m$ and $\tau_S$ and the correlations in their 
measurements they finally quote $\phi_{+-}=43.53\deg \pm0.97\deg$

From a simultaneous fit to the \pic\ and \pio\ data they obtain
$\Delta\phi=\phi_{00}-\phi_{+-}=0.62\deg\pm0.71\deg\pm0.75\deg$, which
combined with the E731 result gives $\Delta\phi=-0.3\deg\pm0.88\deg$.

\subsec{Estimating the error on $\phi_{+-}$}
The E773 estimate of the $\phi_{+-}$ error has been criticized
by Kleinknecht and Luitz (K-L in the following)\rlap.\Ref\kkl{K. 
Kleinknecht and S. Luitz, Phys. Lett. {\bf B336} (1994) 581.}
They quite correctly point out that the
results of an experiment of this kind should be given as
$$\phi_\rho-\phi_{+-}=\f_{\rm meas}\pm\delta\f_{\rm meas}$$
followed by a statement that $\phi_\rho$ is estimated as
$$\phi_\rho=\f_{\rm Est.}\pm\delta\f_{\rm Est.}$$
E773 cannot in fact present their results in such a fashion because of the
analysis method used.
Essentially they use analyticity and the assumption that $|f-\bar
f|\propto p^\beta$, from which it follows that $\f_f=-(1+\beta)\pi/2$.

They then perform grand fits to the data, which span the range
$40<p<160$ Gev/c, floating not only the \K\ parameters of interest but also
the exponent of the power law and the value of $|f-\bar f|$ at 70 GeV/c.

They justify this procedure on the basis of
\pointbegin A fit done in this way properly takes in account all
correlations, assuming of course that $|f-\bar f|\propto p^\beta$, $\f_f=-
(1+\beta)\pi/2$ is correct.
\point They argue that for a large class of functions with small deviations
from a single power law, fitting to a single parameter $\beta_{\rm
Eff.}$, does in fact give a correct answer for the effective (and properly 
weighted) value of $\f_{+-}$.
\point They perform a calculation of the $f-\bar f$ regeneration amplitude
in carbon\rlap,\Ref\wingla{R. A. Briere and B. Winstein, \prl. {\bf75} (1995) 
402.} using Glauber screening and $K-N$ data and find excellent
agreement with data at 3-10 GeV/c\Ref\carit{W. C. Carithers \etal, Nucl.
Phys. {\bf B118} (1978) 333.} and 
40--160 GeV/c\rlap.\refmark{\Schwin}
The total shift between a single power law and the fit using
this procedure is $-0.04\deg$ and they estimate that the ultimate error
on the phase in this type of measurement could be as low as $\pm0.35\deg$.
\endpage

Since they do give the result from the fit for the power, $\beta=-
0.571\pm0.007$, one can reconstruct $\f_f$ and especially the error,
$\delta\f_{\rm Est.}= 90\deg\x0.009=0.63\deg$.
K-L use the complete dispersion relation which I rewrite as
$$\f_f(p_0)=-{\pi\over 2}-\int_0^\infty {\beta\over\pi} {{\rm d}\over p}
\log{\rm coth}|u|^2,\ u=\log({p\over p_0}).$$
If $\beta$ varies slowly with $p$ then
$$\f_f(p_0)=-{\pi\over 2}-\sum\beta_i\int_{p_i}^{p_{i+1}}\cdots=-{\pi\over 
2}-\sum\beta_i I_i$$
and
$$\delta\f_{\rm Est.}=\sum\delta\beta_i\x I_i.$$

A comparison between the error estimates of K-L and E773 is given in 
table 1, 
for $p_0=70$ GeV

\centerline{\tenpoint{\bf Table 1.}}
\vglue 4mm
\begintable
 $p$      |          |\multispan{2}\hfil K-L\hfil  |
\multispan{2}\hfil E773\hfil \nrneg{4pt}
 Interval | Integral | $\delta\beta$ & $\delta\phi$|
$\delta\beta$&  $\delta\phi$ \cr
 0-10     | 0.0912  |    0.025      | 0.002    | 0.007 |    \cr
 10-30    | 0.1876  |    0.025      | 0.005    | 0.007 |    \cr
 30-130   | 0.9368  |    0.020      | 0.019    | 0.007 |    \cr
 130-$\infty$ | 0.3551 | 0.050     | 0.018    | 0.007 |    \cr
\hfill SUM | $\pi/2$ |              | 0.044    |       | 0.011 \cr 
 Error    |          |               | 2.5\deg    |       | 0.63\deg \endtable
By inspection of the data on $|f-\bar f|$, a more reasonable estimate 
of the error, in my opinion, is as given in table 2.

\centerline{\tenpoint{\bf Table 2.}}
\vglue 4mm
\begintable
 Momentum |       |      |     \nrneg{4pt}
 Interval | Integral | $\delta\beta$ | $\delta\phi$ \cr
 0-10     | 0.0912  |    0.020      | 0.0018  \cr
 10-40    | 0.2877  |    0.010      | 0.0029 \cr
 40-160   | 0.9071  |    0.007      | 0.0063 \cr
 160-300  | 0.1354  |    0.014      | 0.0019 \cr
 300-$\infty$ | 0.14938 | 0.03      | 0.0045 \cr
          |            | \hfill | 0.0174 \cr
       |         | \hfill Error | 1.0\deg \endtable

\noindent
The error on the phase gets larger for high and low momenta, which are 
more sensitive to the larger error on $\beta$. The actual momentum 
spectrum  of the data should therefore be used. Using the error in table 
2 gives $\f_{+-}=43.53\deg \pm\ab1.4\deg$.

\section{Combined Results from Different Experiments}
Because of the error estimate uncertainties mentioned
earlier, the correlations between parameters, as well as
between {\it past and new} measurements, it is not wise for me
to try to combine results and get better limits. Better
measurements will come soon, certainly by LP99.

\FIG\dmphi
\REF\cplearc{R. Adler \etal, CPLEAR coll., CERN-PPE/95-112 (1995).}
\REF\dac{R. Carosi \etal, Phys. Let. {\bf B237} (1990) 303.}
\REF\dagz{W. C. Carithers \etal, Phys. Rev. Let. {\bf 34} (1975) 1244.}
\REF\dab{S. Gjesdal \etal, Phys. Let. {\bf B52} (1974) 113.}
\REF\dae{C. Geweniger \etal, Phys. Let. {\bf B52} (1974) 108.}
\REF\daf{C. Geweniger \etal, Phys. Let. {\bf B48} (1974) 487.}
\REF\dad{M. Cullen \etal, Phys. Let. {\bf B32} (1970) 523.}
\REF\dbc{L. K. Gibbons \etal, Phys. Rev. Let. {\bf 70}, 1199
(1993), see also L. K. Gibbons, Thesis, E. Fermi Institute, Univ. of
Chicago (1993).}
The CPLEAR collaboration\refmark{\cplearc} has performed an
analysis for
obtaining the best value for $\Delta m$ and $\phi_{+-}$, taking properly 
into account the fact that different 
experiments have different correlations between the two variables. The 
data\refmark{\cpleara,\cplearb,\Schwin,\dac-\dbc} with their correlations are 
shown in fig. \chapterlabel\dmphi. 

\vbox{\cl{\epsfig{file=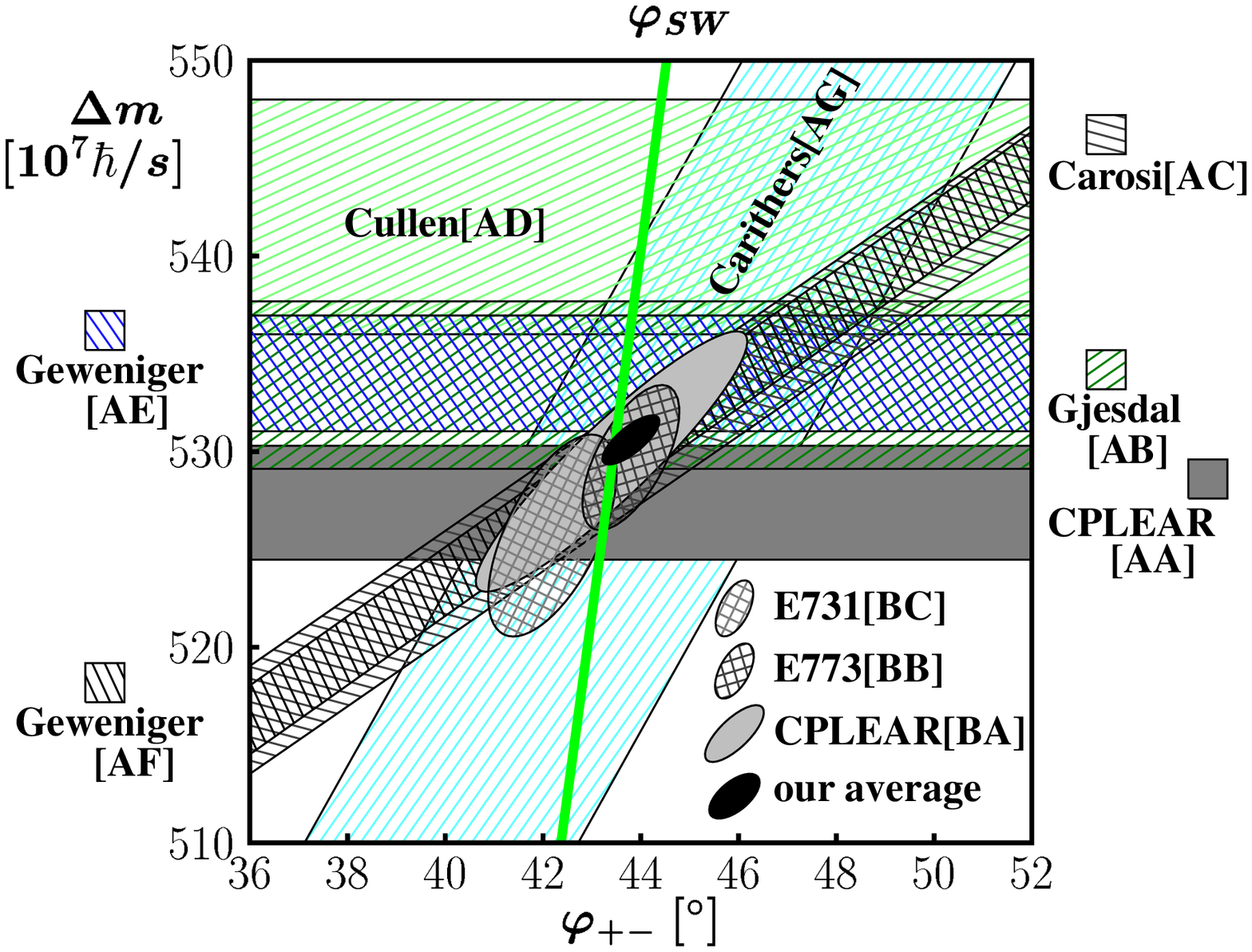,width=12cm}}
\cl{\tenpoint{\bf Fig. \chapterlabel\dmphi.} \ $\Delta m$ and 
$\phi_{\rm SW}$ from ref. \cplearc.}}

\vglue2mm
A maximum likelihood analysis of all data gives 
$\Delta m$=(530.6\plm1.3)\x10\up7 s\up{-1} and 
$\phi_{+-}$=43.75\deg\plm0.6\deg. $\phi_{+-}$ is very close to the 
{\it superweak phase} 
\f\dn{\rm SW}=43.44\deg\plm0.09\deg.           

\subsec{\K\to\pic\gam}
The time dependence of the this decay, like that for two pion case, 
allows extraction of the corresponding parameters:\refmark{\pmgam}
$|\eta_{+-\gam}|=(2.362\pm0.064\pm0.04)\x10^{-3}$ and 
$\phi_{+-\gam}=43.6\deg\pm3.4\deg\pm1.9\deg$.
Comparison with $|\eta_{+-}|\ab|\eps|\ab2.3$, $\phi_{+-}\ab43\deg$ 
gives excellent agreement. This implies that the decay is dominated by 
radiative 
contribution and that all one sees is the $CP$ impurity of the \K\ states.

In fact there is an elegant point to this measurement. Because 
interference is observed (which vanishes between orthogonal states) 
one truly measures the ratio
$$\eta_{+-\gam}={A(\kl\to\pic\gam,\hbox{ \nocp\ }) 
\over A(\ks\to\pic\gam, \hbox{ \C\P\ OK })}$$
which is dominated by E1, inner bremsstrahlung transitions. Thus again one 
is measuring the $CP$ 
impurity of \kl. Direct $CP$ could contribute via E1, direct photon
emission \kl\ decays, 
but it has not been observed, within the sensitivity of the measurement.

\chapter{Tests of $CPT$ Invariance}

The measured parameters in
neutral \K\ decays can be combined to put limits on
possible $CPT$
violation in the \K\ system.
In the following I present a recent analysis of 
Maiani\rlap,\Ref\maia{L. Maiani in {\it The 2\up{nd} DA${\mit\Phi}$NE
Physics Handbook},. eds. L. Maiani, G. Pancheri and 
N. Paver (S.I.S., INFN, 
LNF, Frascati, 1995) 3.}
based on PDG94 data, and also give the limits obtained by CPLEAR and
E773. 

One problem with the neutral \K\ system is that before
arriving to the
answer one has to go through the definition of 21
parameters. Maiani
defines the \ks\ and \kl\ states as
$$\ks\equiv N_S(\sta{K_1}+\eps_S\sta{K_2}),\quad
           \kl\equiv N_L(\sta{K_2}+\eps_L\sta{K_2}),\quad
           \eps_{S,L}\equiv\eps_M\pm\Delta$$
where $\Delta\equiv\delta_{CPT}$ mentioned earlier and the two pion 
amplitudes and $N_S$, $N_L$ are the appropriate state normalization 
coefficients:
$$\eqalign{
A(\ko\to2\pi,I)&\equiv\sqrt{2\over3}(A_I+B_I)e^{i\delta_{I}}\cr
A(\kob\to2\pi, I)&\equiv\sqrt{2\over3}(A^*_I-B^*_I)e^{i\delta_{I}}\cr
}$$
where, for each value of the two pion isospin, $I$=0 and 2, $A$ and $B$ have
the symmetries
$$\vbox{\halign{
\hfil#&\ \hfil#\hfil&\ \hfil#\hfil&\ \hfil#\hfil&\
\hfil#\hfil\cr
       &  $\Re A$    &  $\Im A$    &  $\Re B$    &  $\Im
B$    \cr
 $CP$  &      $+$    &      $-$    &      $-$    &
$+$    \cr
 $T$   &      $+$    &      $-$    &      $+$    &
$+$    \cr
 $CPT$ &      $+$    &      $+$    &      $-$    &
$-$    \cr}
}.$$
The Wu and Yang convention is used, which requires
$A_0$ real and positive.
Then:
$$\eqalign{
\eta_{+-}&=\eps_L+{A(K_2\to\pic)\over
A(K_1\to\pic)}=\eps+\eps'\cr
\eta_{00}&=\eps_L+{A(K_2\to\pio)\over A(K_1\to\pio)}=\eps-
2\eps'\cr
}$$
$$\eps=\eps_M-\Big(\Delta-{\Re B_0\over A_0}\Big),\quad
\eps'=ie^{i(\delta_2-\delta_0)}{\omega\over\sqrt2}
\Big[{\Im A_2\over A_2}-1\Big({\Re B_2\over A_2}-{\Re
B_0\over A_0}
\big)\Big],$$
with $\omega=A_2/A_0$=0.045, from \ks\ and $K^+$ decays.
Four more complex variables need be added for describing the
semileptonic decays, allowing both for \nocpt\ and $\Delta S=-\Delta Q$.
Additional parameters are required for three pion decays. 
\FIG\etafig
The relations above are illustrated in the complex plane in fig. 
\chapterlabel\etafig.
\vglue 1mm
\input etaf

\cl{\tenpoint{\bf Fig. \chapterlabel\etafig} Complex plane 
representation of the \eps\ and $\eta$ parameters, not to scale.}

\section{An Analysis Based on Data from PDG94}

The following data are used by Maiani:
$$\eqalign{
\f_{\rm SW}&\equiv\tan^{-1}{2\Delta m\over \Delta \Gamma}
=43.73\deg\pm0.15\deg\cr
|\eta_{+-}|&=(2.269\pm0.023)\x10^{-3}\cr
|\eta_{00}/\eta_{+-}|&=(0.9955\pm0.0023)\x10^{-3}\cr
\phi_{+-}&=44.3\deg\pm0.8\deg\cr
\phi_{+-}-\phi_{00}&=-1.0\deg\pm1.0\deg\cr
A_L&=(3.27\pm0.12)\x10^{-3}\cr}$$
from which, since $\eps'/\eps$ is so small,
$$\eqalign{|\eps|&=\Big|{2\eta_{+-
}+\eta_{00}\over3}\Big|=
(2.266\pm0.03)\x10^{-10}\cr
\arg(\eps)&=\phi_{+-}+{\phi_{00}-\phi_{+-}\over3}=
44.0\deg\pm1.0\deg\cr}$$
and
$\arg(\eps)-\phi_{\rm SW}=0.3\deg\pm1.0\deg$, 
which implies no $CPT$ violation.
\endpage

From the leptonic asymmetry, $A_L$, Maiani obtains
$$\eqalign{
{\Re B_0\over A_0}&=(-0.1\pm0.6)\x10^{-4}\cr
\Delta&=[(0.0\pm0.6)-i(0.1\pm0.2)]\x10^{-4}\cr}$$
and a limit on the \nocpt\ mass difference
$${M_{11}-M_{22}\over m_K}=(0.0\pm0.9)\x10^{-18}$$

Note that the result above for $\Delta$ is considerably
more stringent
than the direct measurement from CPLEAR, $\Re
\Delta=(0.7\pm5.3\pm4.5)\x10^{-4}$, which however is a direct 
experimental
observation.\Ref\noulis{P. Pavlopolous, to appear in {\it Proc. of the 2\up{nd} 
Workshop on Physics and Detectors for DA${\mit\Phi}$NE}, eds. R. Baldini 
\etal\ (S.I.S., INFN, 
LNF, Frascati, 1995).}

\subsec{Other determinations of the \ko-\kob\ mass difference}
E773, using their values for $\phi_{\rm SW}$, 
$\phi_{+-}$ and $\Delta \phi$
obtain the limit
$$\eqalign{ {|m_{\ko}-m_{\bar K_0}|\over m_{\ko}}&\ab
{2\Delta m\over
m_{\ko}}\, {|\eta_{+-}|\over\sin \phi_{\rm SW}}\x\cr
&|\phi_{+-}-\phi_{\rm SW}+\Delta \phi/3|<1.3\x10^{-
18}\cr}$$

The CPLEAR limit for the mass difference, does
not assume $\Delta S=\Delta Q$ and uses their own new limits
on $\Im x$ and
$\Im \eta_{+-0}$. The limit on \nocpt\ is only slightly
weaker:\refmark{\noulis}
$${|m_{\ko}-m_{\bar K_0}|\over m_{\ko}}<2.2\x10^{-18}$$

\chapter{Rare \K\ Decays}

Rare \K\ decays offer several interesting possibilities, which could 
ultimately open a window beyond the standard model.
They allow the determination of the CKM matrix parameters, as for instance
from the \noc\nop\ decay \kl\to\po$\nu\bar\nu$, as well as from the \C\P\
conserving one \K\up+\to\pip$\nu\bar\nu$.
They also permit the verification of conservation laws which are not strictly
required in the standard model, for instance by searching for \ko\to$\mu e$
decays.

\FIG\etarho
The connection between measurements of neutral \K\ properties and 
branching ratios and
the $\rho$ and $\eta$ parameters of the Wolfenstein parameterization of 
the CKM matrix, ($V_{td}\propto 
1-\eta-i\rho$) is shown {\it schematically} in fig. \chapterlabel\etarho.

\input etarho
\vglue 2mm
\cl{\tenpoint{\bf Fig. \chapterlabel\etarho.} \ Constraints on $\eta$ 
and $\rho$ from physical measurements.}
\vglue2mm

In general the situation valid for the more abundant \K\ decays, \ie\ that
the 
$\nocp\big|_{\rm direct}$ decays have much smaller rates then the 
$\nocp\big|_{\rm indirect}$ ones,
can be reversed for very rare decays. In addition while the evaluation 
of $\eps'$ is particularly unsatisfactory because of the uncertainties 
in the calculation of the
hadronic matrix elements, it is not the case for some rare decays.
A classifications of measurable quantities according to increasing
uncertainties in the calculation of the hadronic matrix elements is
given by Buras\Ref\buras{A. J. Buras, in {\it Phenomenology of 
Unification from Present to Future}, eds. G. Diambrini-Palazzi \etal, 
(World Scientific, 1994) 41.} as:
1. \B(\kl\to\po$\nu\bar\nu$),
2. \B($K^+\to\pi^+\nu\bar\nu$),
3. \B(\kl\to\po\epm), \ \eps\dn{K}, and
4. $\eps'_{K}$, \ \B(\kl\to$\mu\bar\mu]_{\rm SD}$), where SD stands for 
{\it short distance} contributions.
A similar situation holds for the $B$ meson system. The
observation $\eps'\ne0$ remains a unique proof of direct
\nocp. Measurements of 1 through 3, plus present knowledge, over
determine the CKM matrix. So do measurements in the $B$--system.
It would be better to have them both.
Rare \K\ decay experiments are not easy however, just like measuring
\rep\ has not turned out to be. Typical expectations for some of the
interesting decays are:
$$\eqalign{
BR(\kl\to\po\epm,\ \noc\nop]_{dir})&\ab(5\pm2)\x10^{-12}\cr
BR(\kl\to\po\nu\bar\nu)&\ab(3\pm1.2)\x10^{-11}\cr
BR(K^+\to\pi^+\nu\bar\nu)&\ab(1\pm.4)\x10^{-10}\cr
}$$

The most extensive program in this field has been ongoing for a long time at
BNL and I have learned that large statistics have been collected this 
year and are under analysis.
Sensitivities of the order of 10\up{-11} will be reached, although
10\up{-(12{\rm\ or\ }13)} is really necessary.
Experiments with high energy kaon beams have been making excellent progress
toward observing rare decays.

\REF\rarea{M. Weaver \etal, \prl. {\bf72} (1994) 3758.}
\REF\rareb{P. Gu \etal, \prl. {\bf72} (1994) 3000.}
\REF\rarec{D. Roberts \etal, Phys. Rev. {\bf D73} (1994) 1874.}
\REF\rared{T. Nakaya \etal, \prl. {\bf73} (1994) 2169.}
\REF\raree{M. B. Spencer \etal, \prl. {\bf74} (1995) 3323.}
\REF\raref{K. Arisaka \etal, EFI 95-08.}
\REF\rareg{P. Gu \etal, private communication.}

\REF\harea{G. D. Barr \etal, Phys. Lett. {\bf B 304} (1993) 381.}
\REF\hareb{G. D. Barr \etal, Phys. Lett. {\bf B 328} (1994) 528.}
\REF\harec{G. D. Barr \etal, Z. Phys. {\bf C65} (1995) 361.}
\REF\hared{A. Kreutz \etal, Z. Phys. {\bf C65} (1995) 67.}
\REF\haree{K. Kleinknecht, MZ-ETAP/95-4.}
\REF\haref{G. D. Barr \etal, Phys. Lett. {\bf B 351} (1995) 579.}

I will discuss new results from E799-I\refmark{\rarea-\rareg}
(E731 without regenerators) and NA31\rlap.\refmark{\harea-\haref}
The results obtained by the two experiments are 
summarized in the tables below.

\def\less{\ifm{<}}  
\def\pen{\ifm{\pi e\nu}}  \def\Ga{\ifm{\Gamma}}
  
\def\nunu{\ifm{\nu\bar\nu}}  \def\mumu{\ifm{\mu^+\mu^-}}
\def\mepm{\ifm{\mu^\pm e^\mp}}

\vglue4mm
\cl{\tenpoint{\bf Table 3.} E799-I Rare \K-decays Results.}
\vglue3mm
\begintable
  Reaction | Events | \B\ or limit |  Ref.  \cr
 \kl\to\po\nunu |   | \less\pt5.8,-5, | \rarea\   \cr
 \kl\to\epm\epm | 27 | (4.0\plm0.8\plm0.3)\x10\up{-8} | \rareb\   \cr
 \kl\to\pio\gam) |    | \less\pt2.3,-4, |  \rarec\  \cr
 \kl\to\epm\gam\gam, $E_\gamma>$5 MeV | |(6.5\plm1.2\plm0.6)\x10\up{-7} | 
\rared\ \cr
 \kl\to\mumu\gam | 207 | (3.23\plm0.23\plm0.19)\x10\up{-7} | \raree\   \cr
 \kl\to\po\mepm | | \less\pt6.4,-9, | \raref\   \cr
 \kl\to\epm\mumu | 1 | $(2.9^{+6.7}_{-2.4})$\x10\up{-9} | \rareg\ \endtable

\vglue 4mm
\cl{\tenpoint{\bf Table 4.} NA31 Rare \K-decays Results.}
\vglue3mm
\begintable
  Reaction | Events |  \B\ or limit | Ref.   \cr
 \ks\to\po\epm | 0  |  \less\pt1.1,-6, | \harea\   \cr
 \kl\to\pio\gam | 3 |  \less\pt5.6,-6, | \hareb\   \cr
 \kl\to\epm\epm\ | 8 | (10.4\plm3.7\plm1.1)\x10\up{-8} | \harec\   \cr
 \kl\to\pio\po  |   |  0.211\plm0.003 |  \hared\ \cr 
   |\multispan{2} \Ga(\kl\to3\po)/\Ga(\kl\to\pic\po)=1.611\plm0.037 | 
\hared\ \cr
   |\multispan{2} \Ga(\kl\to3\po)/\Ga(\kl\to\pen)=0.545\plm0.01 | \hared\ \cr
 \kl\to\po\gam\gam | 57 | (1.7\plm0.3)\x10\up{-6} | \haree\   \cr
 \kl\to\epm\gam | 2000 | (9.1\plm0.3\plm0.5)\x10\up{-6} | \haree\ \cr
 \kl\to3\gam    |      |  \less\pt2.8,-7, | \haree\ \cr
 \ks\to\gam\gam | 16   | \pt(2.4\plm0.9),-6, | \haref\  \endtable

\noindent
The new results above do not yet determine $\rho$ and $\eta$. 
They do however confirm the feasibility of such program.

\section{Search for \K\up+\to\pip\nunu} This decay, $CP$ allowed, is
best for determining $V_{td}$. At present there is no information, other
than the E787-BNL's limit \B\less\pt7.5,-9,\rlap.\Ref\old{J. F. Haggerty
in {\it Proc. XXVII Int. Conf. on High Energy Physics}, ed. P. J. Bussey
and J. G. Knowles (Institute of Physics Publishing, Bristol 1994)
1341.} The new E787\Ref\new{L. S. Littenberg, private communication.}
detector, which in an engineering run found 12 events of
\K\to$\pi$\mumu, \B\ab10\up{-8}, has collected data for a total of 7500
double density 8 mm tapes. This corresponds to about one
\K\up+\to\pip\nunu\ event. At least 100 are necessary for a first
$V_{td}$ measurements. 

\section{\K\to\gam\gam}
Direct \nocp\ is possible in this channel.
Defining the two photon states, where $L$ and $R$ refer to the photon 
polarizations,
$$\eqalign{
\sta+ &=(\sta{LL}+\sta{RR})/\sqrt2\cr
\sta- &=(\sta{LL}-\sta{RR})/\sqrt2\cr}$$
we have four possibilities for \kl,\ks\to\gam\gam, given below, with the 
expected \B's:
$$\vbox{\halign{\ #\ &\quad#\quad\hfil   &\quad#\hfil\cr
     &\sta+         &\sta-      \cr
\noalign{\vglue 2mm}
\kl  &\pt7,9,, \nocp&\pt6,-4,   \cr
\ks  &\pt2,-6,      &\pt5,-12,, \nocp \cr} }$$

The \nocp\ channels can be isolated by measuring the \gam\ 
polarization, using Dalitz conversion. The present results confirm 
expectations on the \C\P\ conserving channels. Both E799-I and NA31 
have detected \kl\to\epm\epm\ decays, 27 and 8 events 
respectively, finding \B=(3.9\plm0.8, 10\plm4)\x10\up{-8} to be compared 
with the expectation (3.4\plm0.2)\x10\up{-8}. They also have determined 
that $CP\sta{K_2}=-\sta{K_2}$.
NA31 has also observed 69 \K\to\gam\gam\ events, of which 52 are from 
\kl\ and one is background. From this they derive 
\B(\ks\to\gam\gam)=(2.4\plm0.9)\x10\up{-6}. 
These results are in agreement with expectations, still one needs 
sensitivities of 10\up{-12}.

\section{\K\to\mumu}

Second order weak amplitudes give contributions which depend on
$\rho$, with\break
 $BR|_{\rm SD}\ab10^{-9}$. Measurements of the muon polarization 
are necessary. One however 
needs to confirm the calculations for \K\to\gam\gam\to\mumu, which can 
confuse the signal. The following results are relevant
\pointbegin NA31 with 2000 \kl\to\epm\gam\ events finds
\B=(9.1\plm0.3\plm0.5)\x10\up{-6}.
The \B\ depends on the \K\gam\up*\gam\ form factor, with contribution 
from vector meson dominance and the $KK^*\gam$ coupling,
$f(q^2)=f_{VMD}+\alpha_{K^*}f_{KK^*\gam}$. The measured \B\ corresponds
to $\alpha_{K^*} =- 0.27\pm0.1.$ 
\point E799-I observes 207 \kl\to\mumu\gam\ events, giving 
\B=(3.23\plm0.23\plm0.19)\x10\up{-7} and $\alpha_{K^*}=0.13^{+0.21}_{-0.35}$
\point E799-I has found one \kl\to\epm\mumu\ event\rlap,\refmark{\rareg}
on the basis of 
which they estimate the branching ratio as
\B=$(2.9^{+6.7}_{-2.4})$\x10\up{-9}. Expectations are \pt2.3,-9,, from VMD 
and \pt8,-10, for $f(q^2)$=const. Previous limits were \B\less\pt4.9,-6,.

\noindent
At BNL the experiment E871\Ref\twomu{S. G. Wojcicki, private 
communication.}
should have 10\up4 \K\to\mumu\ events 
recorded and, according to the results above,
might extract a first significant value for $\rho$. 

\section{\kl\to\po\epm}

The direct \nocp\ \B\ is expected to be \ab\pt5,-12,.
There are however three contributions to the rate plus a potentially
dangerous background.
\pointbegin \K\dn2\to\po\gam\gam\to\po\epm, a \C\P\ allowed transition.
\point \kl\to\po\epm, from the \kl\ \C\P\ impurity ($\eps\sta{K_1}$). 
\endpage

\point Direct \nocp\ from short distance, second order 
weak contributions, via $s\to d+Z$, the signal of interest.
\point Background from \kl\to\gam\gam\up*\to\epm\gam\to\epm\gam\gam, 
with a photon from final state radiation.

The relevant experimental results are:
\pointbegin NA31: 57 \kl\to\po\gam\gam, 
\B=(1.6\plm0.3)\x10\up{-6}, equivalent to 
\B(\kl\to\po\epm) = \pt5,-13,
\point NA31 finds no \ks\to\po\epm\ events or \B\less\pt1.1,-6,, from 
which\nextline\noindent
\B(\kl\to\po\epm) \ab\ $|\eps|^2(\Gamma_S/\Gamma_L)\B(\ks) < 3.2\x10^{-9}$,
which is not quite good enough yet.
\point 799-I finds 58 \kl\to\epm\gam\gam\ events, 
\B=(6.5\plm1.2\plm0.6)\x10\up{-7}. 

The background from  point 3 above will not be dangerous for the new 
proposed experiments (KTEV and NA48), because of the superior resolution 
of their new electromagnetic
calorimeter.
The observation of direct \nocp\ contributions to \kl\to\po\epm\ 
should be convincing when the necessary sensitivity is reached.

\section{\kl\to\po\nunu}
This process is a pure direct \nocp\ signal. The present limits are far 
from the goal. The 
sensitivities claimed for E799-II and at KEK are around 10\up{-9}. 
Another factor of 100 improvement is necessary.

\baselineskip=13.8pt
\chapter{Other \nocp\ Searches}

Upper limits on the weak $\tau$ electric dipole form factor $\tilde d_W$
have been placed by the LEP experiments.
ALEPH\Ref\alef{D. Buskulic \etal, CERN-PPE/94-175.} finds 
$|\tilde d_W|<1.5\x10^{-17}$,
DELPHI\Ref\delfi{M.-C. Chen \etal, DELPHI 95-113 Phys 548, 1995.} 
gives $|\tilde d_W|<2.1\x10^{-17}$ and
OPAL\Ref\opal{B. Akers \etal, CERN/94-171.} places limits on both the 
real and imaginary parts 
of $\tilde d_W$, $|\Re\tilde d_W|<7.8\x10^{-18}$, $|\Im\tilde d_W|<
4.5\x10^{-17}$. Observation of a non zero value for $\tilde d_W$ is 
proof of direct \nocp.
KEK experiment 246 is approved for a measurement of the muon polarization
in $K^+\to\po\mu^+\nu$, which allows searching for 
\noT\rlap.\Ref\kekmu{J. Imazato, KEK preprint 95-54, May 1995.}
Experiment E871\Ref\eeso{K. B. Luk, private communication.} at FNAL 
will run next spring 
searching for \nocp\ in 
hyperon decay. They will measure the \noc\ asymmetry parameter 
$\alpha$ for $\Lambda$, 
$\overline\Lambda$, $\Xi^-$ and $\overline\Xi^+$ in the decays
$\Xi\to\Lambda\pi$, $\Lambda\to p\pi$ to a sensitivity of
$(\alpha-\bar\alpha)<10^{-4}$. 
A non vanishing value of $\alpha-\bar\alpha$ is unambiguous 
proof of direct \nocp. The expected signal is \pt5,-4,.

\chapter{Future}
Three new experiments: NA48\Ref\epsi{M. Calvetti, to appear in {\it Proc. 
of the 2\up{nd} 
Workshop on Physics and Detectors for DA${\mit\Phi}$NE}, eds. R. Baldini 
\etal\ (S.I.S., INFN, 
LNF, Frascati, 1995).}
in CERN, KTEV\Ref\nktev{B. Winstein, to appear in {\it Proc. of the 2\up{nd} 
Workshop on Physics and Detectors for DA${\mit\Phi}$NE}, eds. R. Baldini 
\etal\ (S.I.S., INFN, LNF, Frascati, 1995).}
at FNAL and KLOE\Ref\kloe{J. Lee-Franzini, to appear in {\it Proc. of 
the 2\up{nd} Workshop on Physics and Detectors for DA${\mit\Phi}$NE}, 
eds. R. Baldini \etal\ (S.I.S., INFN, LNF, Frascati, 1995).} at LNF, are
under construction and will begin taking data in '96 -- '97, with the
primary aim to reach an ultimate error in \rep\ of ${\cal O}$(10\up{-4}).
The sophistication of these experiments takes advantage of our 
experience of two decades of
fixed target and \epm\ collider physics. Fundamental in KLOE
is the possibility of continuous self-calibration while running, via 
processes like Bhabha scattering and charged \K\ decays.

\baselineskip=14pt

\section{NA48}
\FIG\nafe
\vbox{\cl{\epsfig{file=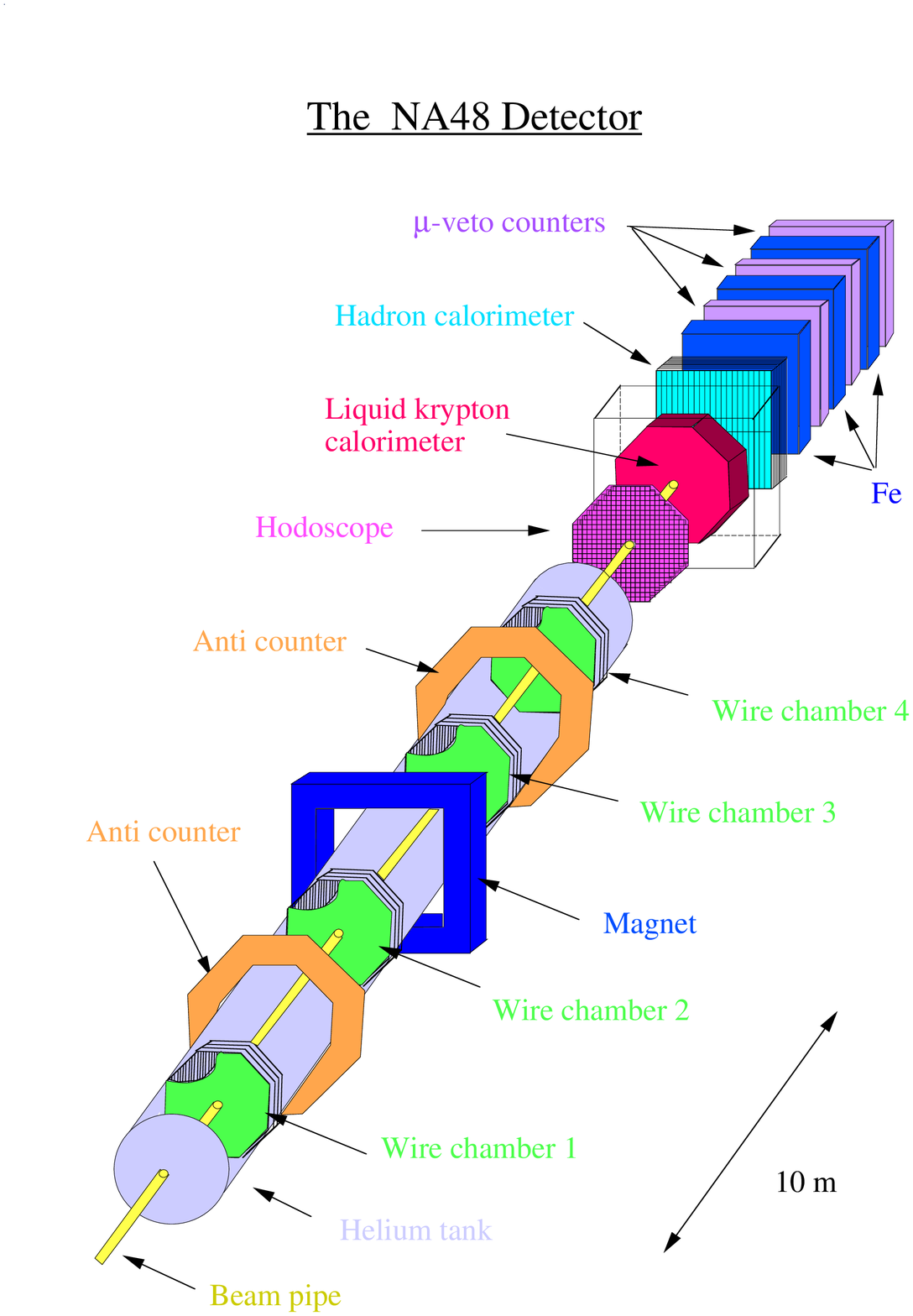,width=13cm}}
\cl{\tenpoint{\bf Fig. \chapterlabel\nafe.} The NA48 experiment at CERN.}}
A new feature of NA48, with respect to its predecessor NA31, is that 
\kl\ and \ks\ 
beams simultaneously illuminated the detector, by the very clever use of 
a bent crystal to deflect a portion of the incident proton beam. This
deflected beam
is brought to a \ks\ production target located close to the 
detector, reducing systematic errors due to different dead times when 
detecting \pic\ or \pio\ \K\ decays. The superior resolution of the 
liquid krypton calorimeter further improves the definition of the 
fiducial regions and improves rejection of 3\po's 
decays. A magnetic spectrometer has also been 
added. Fig. \chapterlabel\nafe\ gives is view of the NA48 setup.

\section{KTEV}
\FIG\ktev
\vbox{\cl{\epsfig{file=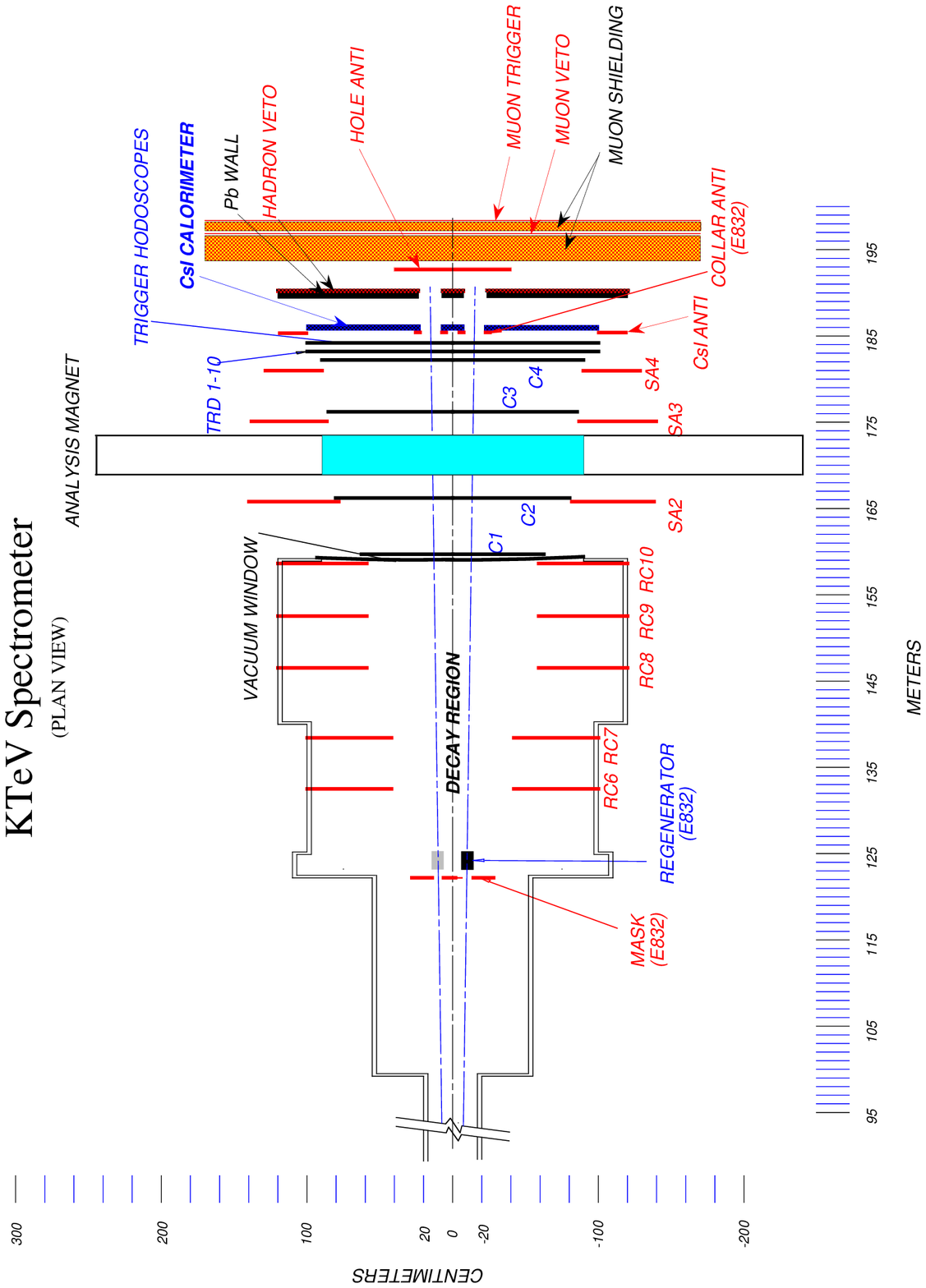,bbllx=1.2cm,bblly=1cm,bburx=16cm,bbury=22.5cm,width=12cm}}
\cl{\tenpoint{\bf Fig. \chapterlabel\ktev.} Plan view of the KTEV 
experiment at FNAL.}}

The KTEV experiment retains the basic principle of E731, with several
significant 
improvements, the most important being the use of CsI crystals for the 
electromagnetic calorimeter. This results in better energy resolution 
which is important for background rejection in the \pio\ channel as well 
as in the search for rare \K\ decays. A plan view of KTEV is shown in 
fig. \chapterlabel\ktev.

\section{KLOE}
\FIG\klosec
The KLOE detector looks very much like a collider detector and will be 
in fact operated at the \DAF\ collider under construction at the 
Laboratori Nazionali di Frascati, LNF. A cross section of KLOE is shown 
in fig. \chapterlabel\klosec.
At \DAF\ \K-meson are produced in 
pairs at rest in the laboratory, via the reaction \epm\to\f\to2\K.
\ab5000 \f-mesons are produced per second 
at a total 
energy of W=1020 MeV and full \DAF\ luminosity. 

\vbox{
\cl{\epsfig{file=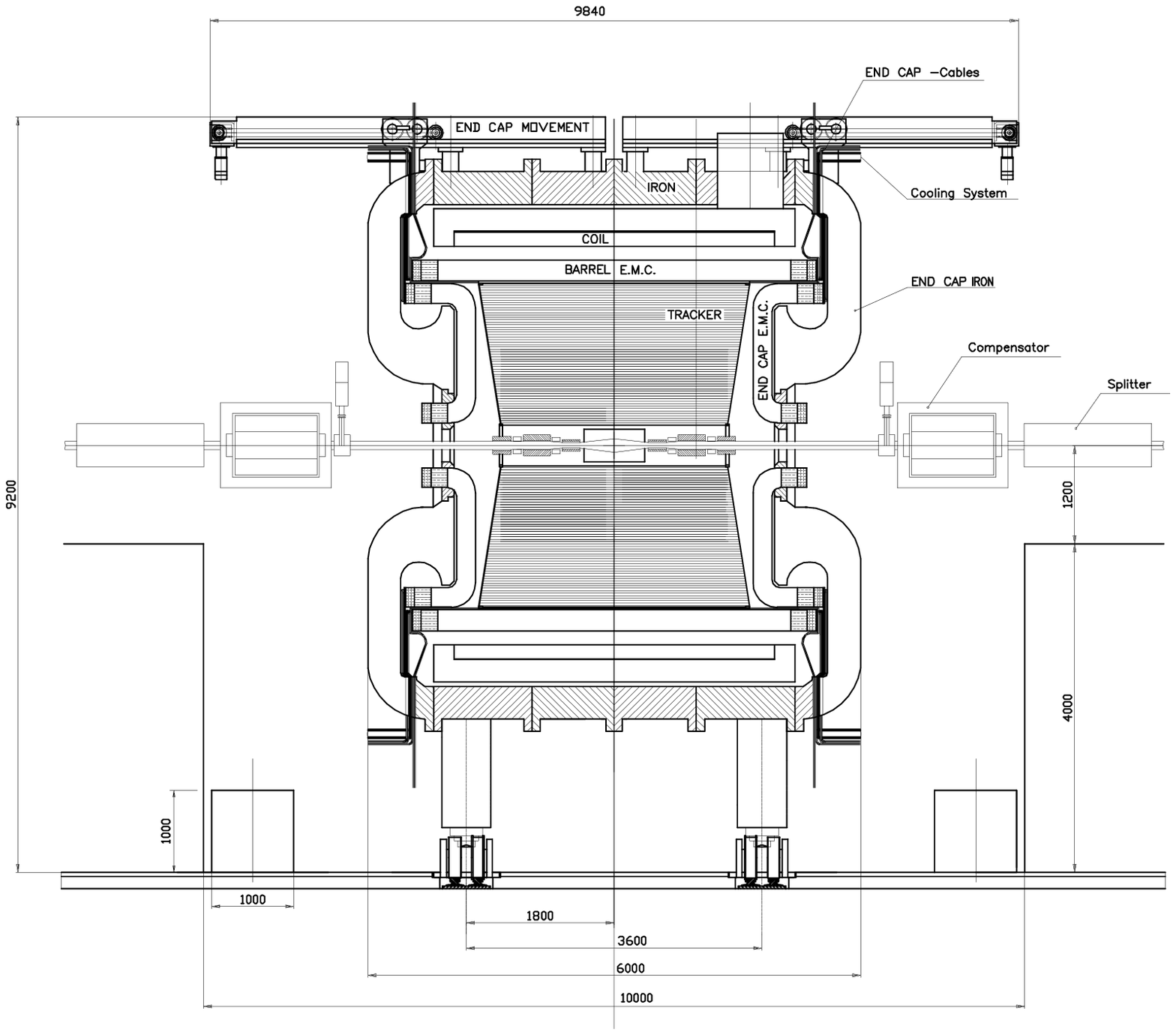,width=15cm}}
\cl{\tenpoint{\bf Fig. \chapterlabel\klosec.} Section of KLOE at \DAF.}}

The two neutral kaons are produced in a pure 
$C$-odd quantum state. This implies that, to a very 
high level of accuracy, the final state is always \ks\kl$-$\kl\ks\ or 
\ko\kob$-$\kob\ko. Tagging of \ks, \kl, \ko, \kob\ is therefore possible.
A pure \ks\ beam of about 10\up{10} per year is a unique possibility at 
\DAF\ at full luminosity. 
The produced kaons are monochromatic, with $\beta$\ab0.2. This allows 
measurement of the flight path of neutral \K's by time of flight. 
\REFS\rosner{I. Duniez, J. Hauser and J. Rosner, Phys. Rev. D {\bf35}, 
2166 (1987).}
\REFSCON\paolo{P. Franzini, in {\it Proc. of the 
Workshop on Physics and Detectors for DA${\mit\Phi}$NE}, ed. G. Pancheri
(S.I.S., INFN, 
LNF, Frascati, 1991) 733.}

Finally 
because of the well defined quantum state, spectacular interference 
effects are observable\rlap,\refsend allowing a totally different 
way of measuring 
\rep, in addition to the classical method of the double ratio ${\cal R}$.
This experiment is however more difficult, because no first order 
cancellations of many systematic errors are possible.

\section{Conclusions}
Ultimately three independent measurements
performed with very different techniques should
be able to determine whether \rep$\ne$0, as long as \rep\ab few\x10\up{-4}.

Each experiment has additional by-products of interest in kaon
physics.
From KTEV and NA48, more precise values of \f\dn{+-} and $\Delta \f$ will
be obtained. KTEV expects to reach an error of 0.5\deg\ in the 
experimental determination
of \f\dn{f} or \f\dn{\rho} using semileptonic decays. NA48 can measure 
\f\dn{+-} by oscillations of the decay rate behind their production targets, 
if $n(\ko)\ne n(\kob)$. The strong correlation between $\Delta m$ and 
\f\dn{+-} does not change. However all errors will be smaller.
Likewise other parameters relevant to testing $CPT$ invariance will be
measured to higher accuracy, {\it e.g.} the charge asymmetry $A_L$ in 
semileptonic decays. In this respect the 
uniqueness of 
\DAF\ is that of providing a tagged, pure \ks\ beam which allows KLOE to 
measure
the charge asymmetry $A_S$ in leptonic decays of \ks-mesons to an accuracy
$\delta A_S$\ab few\x10\up{-4}. The value of $\Gamma_L$ is becoming
relevant in the analysis of the \ko--\kob, \ks--\kl\ system. This is
a measurement which KLOE can perform, improving the accuracy by \ab\x15.

Concerning rare decays the number of events collected by KTEV and NA48 
should increase by a
factor of 100, corresponding to putting limits of few\x10\up{-11} on unobserved
decays and an improvement of a factor ten in the measurable rates.
The statistics available at \DAF\ for \kl\ decays cannot compete with that
of KTEV and NA48. However the tagged \ks\ beam will allow us to improve
the measurements of rare \ks\ decays by three orders of magnitude.

One last open question is a better test of the $\Delta S=\Delta Q$ rule. 
This is
not possible with the \ko-\kob\ state produced at \DAF\ (without 
invoking $CPT$) nor with high energy \K\ beams. \K's tagged via 
strong interactions are required to test the rule.
The copious $K^+K^-$ \ production at \DAF\ provides tagged
\K\up{+}(\K\up{-}) beams which, via charge exchange, results in strangeness
tagged \ko(\kob)'s, much in the same way it is done in CPLEAR. CPLEAR has
collected tens of million events, KLOE can do at least a factor of ten 
better.

A little farther in time, a strong \K\ program at the main injector at FNAL, 
KAMI, if approved, could by the beginning 
of the next millennium, be quite competitive 
with and complementary to the $B$-factories in determining the CKM matrix 
parameters (or finding something wrong with the standard model).

\chapter{Acknowledgements}
I wish to thank P. Pavloupolous and E. Gabathuler for providing me with 
material on the CPLEAR results and L. Maiani for discussions on his 
analysis. I also thank B. Winstein and K. 
Kleinknecht for discussions about the determination of the phase 
\f\dn{+-}. Finally I wish to acknowledge the help of Juliet Lee-Franzini 
for clarifying many points about rare \K-decays and the $CPT$ tests in 
the neutral \K-system.

\immediate\closeout\referencewrite
\referenceopenfalse
\chapter{References}

\input referenc.texauxil

\bye

%% file: phyzelv.tex
%
\catcode`@=11 
%
%
%

\font\fourteenrm=cmr10 scaled\magstep2
\font\twelverm=cmr12
\font\elvrm=cmr10 scaled\magstephalf
\font\ninerm=cmr9            \font\sixrm=cmr6
\font\egtrm=cmr8

\font\fourteenbf=cmbx10 scaled\magstep2
\font\twelvebf=cmbx12
\font\elvbf=cmbx10 scaled\magstephalf
\font\ninebf=cmbx9            \font\sixbf=cmbx6
\font\egtbf=cmbx8
\font\seventeeni=cmmi10 scaled\magstep3     \skewchar\seventeeni='177
\font\fourteeni=cmmi10 scaled\magstep2      \skewchar\fourteeni='177
\font\twelvei=cmmi12                        \skewchar\twelvei='177
\font\elvi=cmmi10 scaled\magstephalf     \skewchar\elvi='177
\font\ninei=cmmi9                           \skewchar\ninei='177
\font\egti=cmmi8                           \skewchar\egti='177
\font\sixi=cmmi6                            \skewchar\sixi='177
\font\seventeensy=cmsy10 scaled\magstep3    \skewchar\seventeensy='60
\font\fourteensy=cmsy10 scaled\magstep2     \skewchar\fourteensy='60
\font\twelvesy=cmsy10 scaled\magstep1       \skewchar\twelvesy='60
\font\elvsy=cmsy10 scaled\magstephalf    \skewchar\elvsy='60
\font\ninesy=cmsy9                          \skewchar\ninesy='60
\font\egtsy=cmsy8                          \skewchar\egtsy='60
\font\sixsy=cmsy6                           \skewchar\sixsy='60

\font\fourteenex=cmex10 scaled\magstep2
\font\twelveex=cmex10 scaled\magstep1

\font\elvex=cmex10 scaled\magstephalf
\font\ninex=cmex10

\font\fourteensl=cmsl10 scaled\magstep2
\font\twelvesl=cmsl12
\font\elvsl=cmsl10 scaled\magstephalf
\font\ninesl=cmsl9

\font\fourteenit=cmti10 scaled\magstep2
\font\twelveit=cmti12
\font\elvit=cmti10 scaled \magstephalf
\font\nineit=cmti9

\font\twelvett=cmtt12
\font\elvtt=cmtt10 scaled \magstephalf
\font\ninett=cmtt9

 \font\twelvecp=cmcsc10 scaled\magstep1
 \font\elvcp=cmcsc10 scaled\magstephalf
 \font\tencp=cmcsc10
 \font\ninecp=cmcsc10
 
 \newfam\cpfam
%
%
\newcount\f@ntkey            \f@ntkey=0
\def\samef@nt{\relax \ifcase\f@ntkey \rm \or\oldstyle \or\or
         \or\it \or\sl \or\bf \or\tt \or\caps \fi }
\def\fourteenpoint{\relax
    \textfont0=\fourteenrm          \scriptfont0=\tenrm
    \scriptscriptfont0=\sevenrm
     \def\rm{\fam0 \fourteenrm \f@ntkey=0 }\relax
    \textfont1=\fourteeni           \scriptfont1=\teni
    \scriptscriptfont1=\seveni
     \def\oldstyle{\fam1 \fourteeni\f@ntkey=1 }\relax
    \textfont2=\fourteensy          \scriptfont2=\tensy
    \scriptscriptfont2=\sevensy
    \textfont3=\fourteenex     \scriptfont3=\fourteenex
    \scriptscriptfont3=\fourteenex
    \def\it{\fam\itfam \fourteenit\f@ntkey=4 }\textfont\itfam=\fourteenit
    \def\sl{\fam\slfam \fourteensl\f@ntkey=5 }\textfont\slfam=\fourteensl
    \scriptfont\slfam=\tensl
    \def\bf{\fam\bffam \fourteenbf\f@ntkey=6 }\textfont\bffam=\fourteenbf
    \scriptfont\bffam=\tenbf     \scriptscriptfont\bffam=\sevenbf
    \def\tt{\fam\ttfam \twelvett \f@ntkey=7 }\textfont\ttfam=\twelvett
    \h@big=11.9\p@{} \h@Big=16.1\p@{} \h@bigg=20.3\p@{} \h@Bigg=24.5\p@{}
    \def\caps{\fam\cpfam \twelvecp \f@ntkey=8 }\textfont\cpfam=\twelvecp
    \setbox\strutbox=\hbox{\vrule height 12pt depth 5pt width\z@}
    \samef@nt}
\def\twelvepoint{\relax
    \textfont0=\twelverm          \scriptfont0=\ninerm
    \scriptscriptfont0=\sixrm
     \def\rm{\fam0 \twelverm \f@ntkey=0 }\relax
    \textfont1=\twelvei           \scriptfont1=\ninei
    \scriptscriptfont1=\sixi
     \def\oldstyle{\fam1 \twelvei\f@ntkey=1 }\relax
    \textfont2=\twelvesy          \scriptfont2=\ninesy
    \scriptscriptfont2=\sixsy
    \textfont3=\twelveex          \scriptfont3=\twelveex
    \scriptscriptfont3=\twelveex
    \def\it{\fam\itfam \twelveit \f@ntkey=4 }\textfont\itfam=\twelveit
    \def\sl{\fam\slfam \twelvesl \f@ntkey=5 }\textfont\slfam=\twelvesl
    \scriptfont\slfam=\ninesl
    \def\bf{\fam\bffam \twelvebf \f@ntkey=6 }\textfont\bffam=\twelvebf
    \scriptfont\bffam=\ninebf     \scriptscriptfont\bffam=\sixbf
    \def\tt{\fam\ttfam \twelvett \f@ntkey=7 }\textfont\ttfam=\twelvett
    \h@big=10.2\p@{}
    \h@Big=13.8\p@{}
    \h@bigg=17.4\p@{}
    \h@Bigg=21.0\p@{}
    \def\caps{\fam\cpfam \twelvecp \f@ntkey=8 }\textfont\cpfam=\twelvecp
    \setbox\strutbox=\hbox{\vrule height 10pt depth 4pt width\z@}
    \samef@nt}
\def\elvpoint{\relax
    \textfont0=\elvrm          \scriptfont0=\egtrm
    \scriptscriptfont0=\sixrm
    \def\rm{\fam0 \elvrm \f@ntkey=0 }\relax
    \textfont1=\elvi           \scriptfont1=\egti
    \scriptscriptfont1=\sixi
    \def\oldstyle{\fam1 \elvi \f@ntkey=1 }\relax
    \textfont2=\elvsy          \scriptfont2=\egtsy
    \scriptscriptfont2=\sixsy
    \textfont3=\elvex          \scriptfont3=\elvex
    \scriptscriptfont3=\elvex
    \def\it{\fam\itfam \elvit \f@ntkey=4 }\textfont\itfam=\elvit
    \def\sl{\fam\slfam \elvsl \f@ntkey=5 }\textfont\slfam=\elvsl
    \def\bf{\fam\bffam \elvbf \f@ntkey=6 }\textfont\bffam=\elvbf
    \scriptfont\bffam=\egtbf     \scriptscriptfont\bffam=\sixbf
    \def\tt{\fam\ttfam \elvtt \f@ntkey=7 }\textfont\ttfam=\elvtt
    \def\caps{\fam\cpfam \elvcp \f@ntkey=8 }\textfont\cpfam=\elvcp
    \setbox\strutbox=\hbox{\vrule height 9.2pt depth 3.7pt width\z@}
    \samef@nt}
\def\tenpoint{\relax
    \textfont0=\tenrm          \scriptfont0=\sevenrm
    \scriptscriptfont0=\fiverm
    \def\rm{\fam0 \tenrm \f@ntkey=0 }\relax
    \textfont1=\teni           \scriptfont1=\seveni
    \scriptscriptfont1=\fivei
    \def\oldstyle{\fam1 \teni \f@ntkey=1 }\relax
    \textfont2=\tensy          \scriptfont2=\sevensy
    \scriptscriptfont2=\fivesy
    \textfont3=\tenex          \scriptfont3=\tenex
    \scriptscriptfont3=\tenex
    \def\it{\fam\itfam \tenit \f@ntkey=4 }\textfont\itfam=\tenit
    \def\sl{\fam\slfam \tensl \f@ntkey=5 }\textfont\slfam=\tensl
    \def\bf{\fam\bffam \tenbf \f@ntkey=6 }\textfont\bffam=\tenbf
    \scriptfont\bffam=\sevenbf     \scriptscriptfont\bffam=\fivebf
    \def\tt{\fam\ttfam \tentt \f@ntkey=7 }\textfont\ttfam=\tentt
    \def\caps{\fam\cpfam \tencp \f@ntkey=8 }\textfont\cpfam=\tencp
    \setbox\strutbox=\hbox{\vrule height 8.5pt depth 3.5pt width\z@}
    \samef@nt}
\def\ninpoint{\relax
    \textfont0=\ninerm          \scriptfont0=\sevenrm
    \scriptscriptfont0=\fiverm
    \def\rm{\fam0 \ninerm \f@ntkey=0 }\relax
    \textfont1=\ninei           \scriptfont1=\seveni
    \scriptscriptfont1=\fivei
    \def\oldstyle{\fam1 \ninei \f@ntkey=1 }\relax
    \textfont2=\ninesy          \scriptfont2=\sevensy
    \scriptscriptfont2=\fivesy
    \textfont3=\ninex          \scriptfont3=\ninex
    \scriptscriptfont3=\ninex
    \def\it{\fam\itfam \nineit \f@ntkey=4 }\textfont\itfam=\nineit
    \def\sl{\fam\slfam \ninesl \f@ntkey=5 }\textfont\slfam=\ninesl
    \def\bf{\fam\bffam \ninebf \f@ntkey=6 }\textfont\bffam=\ninebf
    \scriptfont\bffam=\sevenbf     \scriptscriptfont\bffam=\fivebf
    \def\tt{\fam\ttfam \ninett \f@ntkey=7 }\textfont\ttfam=\ninett
    \def\caps{\fam\cpfam \ninecp \f@ntkey=8 }\textfont\cpfam=\ninecp
    \setbox\strutbox=\hbox{\vrule height 8. pt depth 3.3pt width\z@}
    \samef@nt}
%
%
%
%
\newdimen\h@big  \h@big=8.5\p@
\newdimen\h@Big  \h@Big=11.5\p@
\newdimen\h@bigg  \h@bigg=14.5\p@
\newdimen\h@Bigg  \h@Bigg=17.5\p@
\def\big#1{{\hbox{$\left#1\vbox to\h@big{}\right.\n@space$}}}
\def\Big#1{{\hbox{$\left#1\vbox to\h@Big{}\right.\n@space$}}}
\def\bigg#1{{\hbox{$\left#1\vbox to\h@bigg{}\right.\n@space$}}}
\def\Bigg#1{{\hbox{$\left#1\vbox to\h@Bigg{}\right.\n@space$}}}
%
%
%
\normalbaselineskip = 20pt plus 0.2pt minus 0.1pt
\normallineskip = 1.5pt plus 0.1pt minus 0.1pt
\normallineskiplimit = 1.5pt
\newskip\normaldisplayskip
\normaldisplayskip = 8pt plus 4pt minus 3pt
\newskip\normaldispshortskip
\normaldispshortskip = 6pt plus 2pt
\newskip\normalparskip
\normalparskip = 3pt plus 1pt minus 1pt
\newskip\skipregister
\skipregister = 4pt plus 1pt minus .5pt
\newif\ifsingl@    \newif\ifdoubl@
\newif\iftwelv@    \twelv@true
\def\singlespace{\singl@true\doubl@false\spaces@t}
\def\doublespace{\singl@false\doubl@true\spaces@t}
\def\normalspace{\singl@false\doubl@false\spaces@t}
\def\Elvpoint{\elvpoint\twelv@false\spaces@t}
\def\Tenpoint{\tenpoint\twelv@false\spaces@t}
\def\Ninpoint{\ninpoint\twelv@false\spaces@t}
\def\Twelvepoint{\twelvepoint\twelv@true\spaces@t}
\def\spaces@t{\relax%
 \iftwelv@ \ifsingl@\subspaces@t3:4;\else\subspaces@t1:1;\fi%
 \else \ifsingl@\subspaces@t3:5;\else\subspaces@t4:5;\fi \fi%
 \ifdoubl@ \multiply\baselineskip by 5%
 \divide\baselineskip by 4 \fi \unskip}
\def\subspaces@t#1:#2;{
      \baselineskip = \normalbaselineskip
      \multiply\baselineskip by #1 \divide\baselineskip by #2
      \lineskip = \normallineskip
      \multiply\lineskip by #1 \divide\lineskip by #2
      \lineskiplimit = \normallineskiplimit
      \multiply\lineskiplimit by #1 \divide\lineskiplimit by #2
      \parskip = \normalparskip
      \multiply\parskip by #1 \divide\parskip by #2
      \abovedisplayskip = \normaldisplayskip
      \multiply\abovedisplayskip by #1 \divide\abovedisplayskip by #2
      \belowdisplayskip = \abovedisplayskip
      \abovedisplayshortskip = \normaldispshortskip
      \multiply\abovedisplayshortskip by #1
        \divide\abovedisplayshortskip by #2
      \belowdisplayshortskip = \abovedisplayshortskip
      \advance\belowdisplayshortskip by \belowdisplayskip
      \divide\belowdisplayshortskip by 2
      \smallskipamount = \skipregister
      \multiply\smallskipamount by #1 \divide\smallskipamount by #2
      \medskipamount = \smallskipamount \multiply\medskipamount by 2
      \bigskipamount = \smallskipamount \multiply\bigskipamount by 4 }
\def\normalbaselines{ \baselineskip=\normalbaselineskip
   \lineskip=\normallineskip \lineskiplimit=\normallineskip
   \iftwelv@\else \multiply\baselineskip by 4 \divide\baselineskip by 5
     \multiply\lineskiplimit by 4 \divide\lineskiplimit by 5
     \multiply\lineskip by 4 \divide\lineskip by 5 \fi }
\Twelvepoint  
\interlinepenalty=50
\interfootnotelinepenalty=5000
\predisplaypenalty=9000
\postdisplaypenalty=500
\hfuzz=1pt
\vfuzz=0.2pt
%
%
%
\def\pagecontents{
   \ifvoid\topins\else\unvbox\topins\vskip\skip\topins\fi
   \dimen@ = \dp255 \unvbox255
   \ifvoid\footins\else\vskip\skip\footins\footrule\unvbox\footins\fi
   \ifr@ggedbottom \kern-\dimen@ \vfil \fi }
\def\makeheadline{\vbox to 0pt{ \skip@=\topskip
      \advance\skip@ by -12pt \advance\skip@ by -2\normalbaselineskip
      \vskip\skip@ \line{\vbox to 12pt{}\the\headline} \vss
      }\nointerlineskip}
\def\makefootline{\baselineskip = 1.5\normalbaselineskip
                 \line{\the\footline}}
\newif\iffrontpage
\newif\ifletterstyle
\newif\ifp@genum
\def\nopagenumbers{\p@genumfalse}
\def\pagenumbers{\p@genumtrue}
\pagenumbers
\newtoks\paperheadline
\newtoks\letterheadline
\newtoks\letterfrontheadline
\newtoks\lettermainheadline
\newtoks\paperfootline
\newtoks\letterfootline
\newtoks\date
\footline={\ifletterstyle\the\letterfootline\else\the\paperfootline\fi}
\paperfootline={\hss\iffrontpage\else\ifp@genum\tenrm\folio\hss\fi\fi}
\letterfootline={\hfil}
\headline={\ifletterstyle\the\letterheadline\else\the\paperheadline\fi}
\paperheadline={\hfil}
\letterheadline{\iffrontpage\the\letterfrontheadline
     \else\the\lettermainheadline\fi}
\lettermainheadline={\rm\ifp@genum page \ \folio\fi\hfil\the\date}
\def\monthname{\relax\ifcase\month 0/\or January\or February\or
   March\or April\or May\or June\or July\or August\or September\or
   October\or November\or December\else\number\month/\fi}
\date={\monthname\ \number\day, \number\year}
\countdef\pagenumber=1  \pagenumber=1
\def\advancepageno{\global\advance\pageno by 1
   \ifnum\pagenumber<0 \global\advance\pagenumber by -1
    \else\global\advance\pagenumber by 1 \fi \global\frontpagefalse }
\def\folio{\ifnum\pagenumber<0 \romannumeral-\pagenumber
           \else \number\pagenumber \fi }
\def\footrule{\dimen@=\prevdepth\nointerlineskip
   \vbox to 0pt{\vskip -0.25\baselineskip \hrule width 0.35\hsize \vss}
   \prevdepth=\dimen@ }
\newtoks\foottokens
\foottokens={\Tenpoint\singlespace}
\newdimen\footindent
\footindent=24pt
\def\vfootnote#1{\insert\footins\bgroup  \the\foottokens
   \interlinepenalty=\interfootnotelinepenalty \floatingpenalty=20000
   \splittopskip=\ht\strutbox \boxmaxdepth=\dp\strutbox
   \leftskip=\footindent \rightskip=\z@skip
   \parindent=0.5\footindent \parfillskip=0pt plus 1fil
   \spaceskip=\z@skip \xspaceskip=\z@skip
   \Textindent{$ #1 $}\footstrut\futurelet\next\fo@t}
\def\Textindent#1{\noindent\llap{#1\enspace}\ignorespaces}
\def\footnote#1{\attach{#1}\vfootnote{#1}}

\def\foot{\attach\footsymbolgen\vfootnote{\footsymbol}}
\let\footsymbol=\star
\newcount\lastf@@t           \lastf@@t=-1
\newcount\footsymbolcount    \footsymbolcount=0
\newif\ifPhysRev
\def\footsymbolgen{\relax \ifPhysRev \iffrontpage \NPsymbolgen\else
      \PRsymbolgen\fi \else \NPsymbolgen\fi
   \global\lastf@@t=\pageno \footsymbol }
\def\NPsymbolgen{\ifnum\footsymbolcount<0 \global\footsymbolcount=0\fi
   {\iffrontpage \else \advance\lastf@@t by 1 \fi
    \ifnum\lastf@@t<\pageno \global\footsymbolcount=0
     \else \global\advance\footsymbolcount by 1 \fi }
   \ifcase\footsymbolcount \fd@f\star\or \fd@f\dagger\or \fd@f\ast\or
    \fd@f\ddagger\or \fd@f\natural\or \fd@f\diamond\or \fd@f\bullet\or
    \fd@f\nabla\else \fd@f\dagger\global\footsymbolcount=0 \fi }
\def\fd@f#1{\xdef\footsymbol{#1}}
\def\PRsymbolgen{\ifnum\footsymbolcount>0 \global\footsymbolcount=0\fi
      \global\advance\footsymbolcount by -1
      \xdef\footsymbol{\sharp\number-\footsymbolcount} }
\def\space@ver#1{\let\@sf=\empty \ifmmode #1\else \ifhmode
   \edef\@sf{\spacefactor=\the\spacefactor}\unskip${}#1$\relax\fi\fi}
\def\attach#1{\space@ver{\strut^{\mkern 2mu #1} }\@sf\ }
%
%
\def\smallsize{\relax
\font\eightrm=cmr8
\font\eightbf=cmbx8
\font\eighti=cmmi8
\font\eightsy=cmsy8
\font\eightsl=cmsl8
\font\eightit=cmti8
\font\eightt=cmtt8
\def\eightpoint{\relax
\textfont0=\eightrm  \scriptfont0=\sixrm
\scriptscriptfont0=\sixrm
\def\rm{\fam0 \eightrm \f@ntkey=0}\relax
\textfont1=\eighti  \scriptfont1=\sixi
\scriptscriptfont1=\sixi
\def\oldstyle{\fam1 \eighti \f@ntkey=1}\relax
\textfont2=\eightsy  \scriptfont2=\sixsy
\scriptscriptfont2=\sixsy
\textfont3=\tenex  \scriptfont3=\tenex
\scriptscriptfont3=\tenex
\def\it{\fam\itfam \eightit \f@ntkey=4 }\textfont\itfam=\eightit
\def\sl{\fam\slfam \eightsl \f@ntkey=5 }\textfont\slfam=\eightsl
\def\bf{\fam\bffam \eightbf \f@ntkey=6 }\textfont\bffam=\eightbf
\scriptfont\bffam=\sixbf   \scriptscriptfont\bffam=\sixbf
\def\tt{\fam\ttfam \eightt \f@ntkey=7 }
\def\caps{\fam\cpfam \tencp \f@ntkey=8 }\textfont\cpfam=\tencp
\setbox\strutbox=\hbox{\vrule height 7.35pt depth 3.02pt width\z@}
\samef@nt}
\def\Eightpoint{\eightpoint \relax
  \ifsingl@\subspaces@t2:5;\else\subspaces@t3:5;\fi
  \ifdoubl@ \multiply\baselineskip by 5
            \divide\baselineskip by 4\fi }
\parindent=16.67pt
\itemsize=25pt
\thinmuskip=2.5mu
\medmuskip=3.33mu plus 1.67mu minus 3.33mu
\thickmuskip=4.17mu plus 4.17mu
\def\thinspace{\kern .13889em }
\def\negthinspace{\kern-.13889em }
\def\enspace{\kern.416667em }
 
\def\enskip{\hskip.416667em\relax}
\def\quad{\hskip.83333em\relax}
\def\qquad{\hskip1.66667em\relax}
\def\crr{\cropen{8.3333pt}}
\foottokens={\Eightpoint\singlespace}
\def\papersize{\vsize=38.67pc\hsize=29.17pc\hoffset=3.44pc\voffset=3.7pc
               \skip\footins=\bigskipamount}
\def\lettersize{\hsize=5.417in\vsize=7.08in\hoffset=0in\voffset=.834in
   \skip\footins=\smallskipamount \multiply\skip\footins by 3 }
\def\attach##1{\space@ver{\strut^{\mkern 1.6667mu ##1} }\@sf\ }
\def\PH@SR@V{\doubl@true\baselineskip=20.08pt plus .1667pt minus .0833pt
             \parskip = 2.5pt plus 1.6667pt minus .8333pt }
\def\author##1{\vskip\frontpageskip\titlestyle{\tencp ##1}\nobreak}
\def\address##1{\par\kern 4.16667pt\titlestyle{\tenpoint\it ##1}}
\def\andaddress{\par\kern 4.16667pt \centerline{\sl and} \address}
\def\SLAC{\address{Stanford Linear Accelerator Center\break
      Stanford University, Stanford, California, 94305}}
\def\abstract{\vskip\frontpageskip\centerline{\twelverm ABSTRACT}
              \vskip\headskip }
\def\submit##1{\par\nobreak\vfil\nobreak\medskip
   \centerline{Submitted to \sl ##1}}
\def\doeack{\foot{Work supported by the Department of Energy,
      contract $\caps DE-AC03-76SF00515$.}}
\def\cases##1{\left\{\,\vcenter{\Tenpoint\m@th
    \ialign{$####\hfil$&\quad####\hfil\crcr##1\crcr}}\right.}
\def\matrix##1{\,\vcenter{\Tenpoint\m@th
    \ialign{\hfil$####$\hfil&&\quad\hfil$####$\hfil\crcr
      \mathstrut\crcr\noalign{\kern-\baselineskip}
     ##1\crcr\mathstrut\crcr\noalign{\kern-\baselineskip}}}\,}
\Tenpoint \paperstyle
}
%
%
%
\newcount\chapternumber      \chapternumber=0
\newcount\sectionnumber      \sectionnumber=0
\newcount\equanumber         \equanumber=0
\let\chapterlabel=0
\newtoks\chapterstyle        \chapterstyle={\Number}
\newskip\chapterskip         \chapterskip=\bigskipamount
\newskip\sectionskip         \sectionskip=\medskipamount
\newskip\headskip            \headskip=6pt plus 1pt minus 3pt
\newdimen\chapterminspace    \chapterminspace=5pc
\newdimen\sectionminspace    \sectionminspace=3pc
\newdimen\referenceminspace  \referenceminspace=5pc
\def\chapterreset{\global\advance\chapternumber by 1
   \ifnum\the\equanumber<0 \else\global\equanumber=0\fi
   \figurecount=0  \tablecount=0
   \sectionnumber=0 \makel@bel
   }
\def\makel@bel{\xdef\chapterlabel{%
\the\chapterstyle{\the\chapternumber}.}}
\def\sectionlabel{\number\sectionnumber \quad }
\def\alphabetic#1{\count255='140 \advance\count255 by #1\char\count255}
\def\Alphabetic#1{\count255='100 \advance\count255 by #1\char\count255}
\def\Roman#1{\uppercase\expandafter{\romannumeral #1}}
\def\roman#1{\romannumeral #1}
\def\Number#1{\number #1}
\def\unnumberedchapters{\let\makel@bel=\relax \let\chapterlabel=\relax
\let\sectionlabel=\relax \equanumber=-1 }
\def\titlestyle#1{\par\begingroup \interlinepenalty=9999
     \leftskip=0.02\hsize plus 0.23\hsize minus 0.02\hsize
     \rightskip=\leftskip \parfillskip=0pt
     \hyphenpenalty=9000 \exhyphenpenalty=9000
     \tolerance=9999 \pretolerance=9000
     \spaceskip=0.333em \xspaceskip=0.5em
     \iftwelv@\fourteenpoint\else\twelvepoint\fi
   \noindent #1\par\endgroup }
\def\spacecheck#1{\dimen@=\pagegoal\advance\dimen@ by -\pagetotal
   \ifdim\dimen@<#1 \ifdim\dimen@>0pt \vfil\break \fi\fi}
\def\chapter#1{\par \penalty-300 \vskip\chapterskip
   \spacecheck\chapterminspace
   \chapterreset \titlestyle{\chapterlabel \ #1}
   \nobreak\vskip\headskip \penalty 30000
   \wlog{\string\chapter\ \chapterlabel} }

\def\section#1{\par \ifnum\the\lastpenalty=30000\else
   \penalty-200\vskip\sectionskip \spacecheck\sectionminspace\fi
   \wlog{\string\section\ \chapterlabel \the\sectionnumber}
   \global\advance\sectionnumber by 1  \noindent
   {\caps\enspace\chapterlabel \sectionlabel #1}\par
   \nobreak\vskip\headskip \penalty 30000 }
\def\subsection#1{\par
   \ifnum\the\lastpenalty=30000\else \penalty-100\smallskip \fi
   \noindent\undertext{#1}\enspace \vadjust{\penalty5000}}
\let\subsec=\subsection
\def\undertext#1{\vtop{\hbox{#1}\kern 1pt \hrule}}
\def\APPENDIX#1#2{\par\penalty-300\vskip\chapterskip
   \spacecheck\chapterminspace \chapterreset \xdef\chapterlabel{#1}
   \titlestyle{APPENDIX #2} \nobreak\vskip\headskip \penalty 30000
   \wlog{\string\Appendix\ \chapterlabel} }
\def\Appendix#1{\APPENDIX{#1}{#1}}
\def\appendix{\APPENDIX{A}{}}
%
%
%
\def\eqname#1{\relax \ifnum\the\equanumber<0%
     \xdef#1{{\noexpand\rm(\number-\equanumber)}}%
     \global\advance\equanumber by -1%
    \else \global\advance\equanumber by 1%
      \xdef#1{{\noexpand\rm(\chapterlabel \number\equanumber)}} \fi}
\def\eqinsert#1{\noalign{\dimen@=\prevdepth \nointerlineskip
   \setbox0=\hbox to\displaywidth{\hfil #1}
   \vbox to 0pt{\vss\hbox{$\!\box0\!$}\kern-0.5\baselineskip}
   \prevdepth=\dimen@}}
%

%

%

%
%
\def\GENITEM#1;#2{\par \hangafter=0 \hangindent=#1
    \Textindent{$ #2 $}\ignorespaces}
\outer\def\newitem#1=#2;{\gdef#1{\GENITEM #2;}}
\newdimen\itemsize                \itemsize=30pt
\newitem\item=1\itemsize;
\newitem\sitem=1.75\itemsize;     
\newitem\ssitem=2.5\itemsize;     
\outer\def\newlist#1=#2&#3&#4;{\toks0={#2}\toks1={#3}%
   \count255=\escapechar \escapechar=-1
   \alloc@0\list\countdef\insc@unt\listcount     \listcount=0
   \edef#1{\par
      \countdef\listcount=\the\allocationnumber
      \advance\listcount by 1
      \hangafter=0 \hangindent=#4
      \Textindent{\the\toks0{\listcount}\the\toks1}}
   \expandafter\expandafter\expandafter
    \edef\c@t#1{begin}{\par
      \countdef\listcount=\the\allocationnumber \listcount=1
      \hangafter=0 \hangindent=#4
      \Textindent{\the\toks0{\listcount}\the\toks1}}
   \expandafter\expandafter\expandafter
    \edef\c@t#1{con}{\par \hangafter=0 \hangindent=#4 \noindent}
   \escapechar=\count255}
\def\c@t#1#2{\csname\string#1#2\endcsname}
\newlist\point=\Number&.&1.0\itemsize;
\newlist\subpoint=(\alphabetic&)&1.75\itemsize;
\newlist\subsubpoint=(\roman&)&2.5\itemsize;
%

%
%
%
\newcount\referencecount     \referencecount=0
\newif\ifreferenceopen       \newwrite\referencewrite
\newtoks\rw@toks
\def\NPrefmark#1{\attach{\scriptstyle [ #1 ] }}
\let\PRrefmark=\attach
\def\refmark#1{\relax\ifPhysRev\PRrefmark{#1}\else\NPrefmark{#1}\fi}
\def\refend{\refmark{\number\referencecount}}
\newcount\lastrefsbegincount \lastrefsbegincount=0
\def\refsend{\refmark{\count255=\referencecount
   \advance\count255 by-\lastrefsbegincount
   \ifcase\count255 \number\referencecount
   \or \number\lastrefsbegincount,\number\referencecount
   \else \number\lastrefsbegincount-\number\referencecount \fi}}
\def\refch@ck{\chardef\rw@write=\referencewrite
   \ifreferenceopen \else \referenceopentrue
   \immediate\openout\referencewrite=referenc.texauxil \fi}
%
{\catcode`\^^M=\active 
  \gdef\obeyendofline{\catcode`\^^M\active \let^^M\ }}%
%
{\catcode`\^^M=\active 
  \gdef\ignoreendofline{\catcode`\^^M=5}}
{\obeyendofline\gdef\rw@start#1{\def\t@st{#1} \ifx\t@st\blankend%
\endgroup \@sf \relax \else \ifx\t@st\bl@nkend \endgroup \@sf \relax%
\else \rw@begin#1
\backtotext
\fi \fi } }
{\obeyendofline\gdef\rw@begin#1
{\def\n@xt{#1}\rw@toks={#1}\relax%
\rw@next}}
\def\blankend{}
{\obeylines\gdef\bl@nkend{
}}
\newif\iffirstrefline  \firstreflinetrue
\def\rwr@teswitch{\ifx\n@xt\blankend \let\n@xt=\rw@begin %
 \else\iffirstrefline \global\firstreflinefalse%
\immediate\write\rw@write{\noexpand\obeyendofline \the\rw@toks}%
\let\n@xt=\rw@begin%
      \else\ifx\n@xt\rw@@d \def\n@xt{\immediate\write\rw@write{%
        \noexpand\ignoreendofline}\endgroup \@sf}%
             \else \immediate\write\rw@write{\the\rw@toks}%
             \let\n@xt=\rw@begin\fi\fi \fi}
\def\rw@next{\rwr@teswitch\n@xt}
\def\rw@@d{\backtotext} \let\rw@end=\relax
\let\backtotext=\relax

\newdimen\refindent     \refindent=30pt
\def\refitem#1{\par \hangafter=0 \hangindent=\refindent \Textindent{#1}}
\def\REFNUM#1{\space@ver{}\refch@ck \firstreflinetrue%
 \global\advance\referencecount by 1 \xdef#1{\the\referencecount}}
\def\refnum#1{\space@ver{}\refch@ck \firstreflinetrue%
 \global\advance\referencecount by 1 \xdef#1{\the\referencecount}\refend}

\def\REF#1{\REFNUM#1%
 \immediate\write\referencewrite{%
 \noexpand\refitem{#1.}}%
\begingroup\obeyendofline\rw@start}
\def\ref{\refnum\?%
 \immediate\write\referencewrite{\noexpand\refitem{\?.}}%
\begingroup\obeyendofline\rw@start}
\def\Ref#1{\refnum#1%
 \immediate\write\referencewrite{\noexpand\refitem{#1.}}%
\begingroup\obeyendofline\rw@start}
\def\REFS#1{\REFNUM#1\global\lastrefsbegincount=\referencecount
\immediate\write\referencewrite{\noexpand\refitem{#1.}}%
\begingroup\obeyendofline\rw@start}

\def\REFSCON#1{\REF#1}
\newcount\figurecount     \figurecount=0
\newif\iffigureopen       \newwrite\figurewrite
\def\figch@ck{\chardef\rw@write=\figurewrite \iffigureopen\else
   \immediate\openout\figurewrite=figures.texauxil
   \figureopentrue\fi}
\def\FIGNUM#1{\space@ver{}\figch@ck \firstreflinetrue%
 \global\advance\figurecount by 1 \xdef#1{\the\figurecount}}
\def\FIG#1{\FIGNUM#1
   \immediate\write\figurewrite{\noexpand\refitem{#1.}}%
   \begingroup\obeyendofline\rw@start}
\def\fig{\FIGNUM\? fig.~\?%
\immediate\write\figurewrite{\noexpand\refitem{\?.}}%
\begingroup\obeyendofline\rw@start}
\def\figure{\FIGNUM\? figure~\?
   \immediate\write\figurewrite{\noexpand\refitem{\?.}}%
   \begingroup\obeyendofline\rw@start}
\def\Fig{\FIGNUM\? Fig.~\?%
\immediate\write\figurewrite{\noexpand\refitem{\?.}}%
\begingroup\obeyendofline\rw@start}
\def\Figure{\FIGNUM\? Figure~\?%
\immediate\write\figurewrite{\noexpand\refitem{\?.}}%
\begingroup\obeyendofline\rw@start}
\newcount\tablecount     \tablecount=0
\newif\iftableopen       \newwrite\tablewrite
\def\tabch@ck{\chardef\rw@write=\tablewrite \iftableopen\else
   \immediate\openout\tablewrite=tables.texauxil
   \tableopentrue\fi}
\def\TABNUM#1{\space@ver{}\tabch@ck \firstreflinetrue%
 \global\advance\tablecount by 1 \xdef#1{\the\tablecount}}
\def\TABLE#1{\TABNUM#1
   \immediate\write\tablewrite{\noexpand\refitem{#1.}}%
   \begingroup\obeyendofline\rw@start}
\def\Table{\TABNUM\? Table~\?%
\immediate\write\tablewrite{\noexpand\refitem{\?.}}%
\begingroup\obeyendofline\rw@start}
%

%
%
%
\def\masterreset{\global\pagenumber=1 \global\chapternumber=0
   \ifnum\the\equanumber<0\else \global\equanumber=0\fi
   \global\sectionnumber=0
   \global\referencecount=0 \global\figurecount=0 \global\tablecount=0 }
\def\FRONTPAGE{\ifvoid255\else\vfill\penalty-2000\fi
      \masterreset\global\frontpagetrue
      \global\lastf@@t=0 \global\footsymbolcount=0}
\let\Frontpage=\FRONTPAGE
\def\paperstyle{\letterstylefalse\normalspace\papersize}
\def\letterstyle{\letterstyletrue\singlespace\lettersize}
\def\papersize{\hsize=161mm\vsize=240mm\hoffset=0mm\voffset=0mm
               \skip\footins=\bigskipamount}
\def\lettersize{\hsize=161mm\vsize=247mm\hoffset=0mm\voffset=mm
   \skip\footins=\smallskipamount \multiply\skip\footins by 3 }
\paperstyle   
%
%
\def\MEMO{\letterstyle\FRONTPAGE \letterfrontheadline={\hfil}
    \line{\quad\fourteenrm SLAC MEMORANDUM\hfil\twelverm\the\date\quad}
    \medskip \memod@f}

\def\memit@m#1{\smallskip \hangafter=0 \hangindent=1in
      \Textindent{\caps #1}}
\def\memod@f{\xdef\to{\memit@m{To:}}\xdef\from{\memit@m{From:}}%
     \xdef\topic{\memit@m{Topic:}}\xdef\subject{\memit@m{Subject:}}%
     \xdef\rule{\bigskip\hrule height 1pt\bigskip}}
\memod@f
\newskip\lettertopfil
\lettertopfil = 0pt plus 1.5in minus 0pt
\newskip\letterbottomfil
\letterbottomfil = 0pt plus 2.3in minus 0pt
\newskip\spskip \setbox0\hbox{\ } \spskip=-1\wd0
\def\addressee#1{\medskip\rightline{\the\date\hskip 30pt} \bigskip
   \vskip\lettertopfil
   \ialign to\hsize{\strut ##\hfil\tabskip 0pt plus \hsize \cr #1\crcr}
   \medskip\noindent\hskip\spskip}
\newskip\signatureskip       \signatureskip=40pt
\def\signed#1{\par \penalty 9000 \bigskip \dt@pfalse
  \everycr={\noalign{\ifdt@p\vskip\signatureskip\global\dt@pfalse\fi}}
  \setbox0=\vbox{\singlespace \halign{\tabskip 0pt \strut ##\hfil\cr
   \noalign{\global\dt@ptrue}#1\crcr}}
  \line{\hskip 0.5\hsize minus 0.5\hsize \box0\hfil} \medskip }

\def\endletter{\ifnum\pagenumber=1 \vskip\letterbottomfil\supereject
\else \vfil\supereject \fi}
\newbox\letterb@x
\def\lettertext{\par\unvcopy\letterb@x\par}
\def\multiletter{\setbox\letterb@x=\vbox\bgroup
      \everypar{\vrule height 1\baselineskip depth 0pt width 0pt }
      \singlespace \topskip=\baselineskip }
\def\letterend{\par\egroup}
%
%
%
\newskip\frontpageskip
\newtoks\pubtype
\newtoks\Pubnum
\newtoks\pubnum
\newif\ifp@bblock  \p@bblocktrue
\def\PH@SR@V{\doubl@true \baselineskip=24.1pt plus 0.2pt minus 0.1pt
             \parskip= 3pt plus 2pt minus 1pt }
\def\PHYSREV{\paperstyle\PhysRevtrue\PH@SR@V}
\def\titlepage{\FRONTPAGE\paperstyle\ifPhysRev\PH@SR@V\fi
   \ifp@bblock\p@bblock\fi}
\def\nopubblock{\p@bblockfalse}
\def\endpage{\vfil\break}
\frontpageskip=1\medskipamount plus .5fil
\pubtype={\tensl Preliminary Version}
\Pubnum={$\caps SLAC - PUB - \the\pubnum $}
\pubnum={0000}
\def\p@bblock{\begingroup \tabskip=\hsize minus \hsize
   \baselineskip=1.5\ht\strutbox \topspace-2\baselineskip
   \halign to\hsize{\strut ##\hfil\tabskip=0pt\crcr
   \the\Pubnum\cr \the\date\cr \the\pubtype\cr}\endgroup}
\def\title#1{\vskip\frontpageskip \titlestyle{#1} \vskip\headskip }
\def\author#1{\vskip\frontpageskip\titlestyle{\twelvecp #1}\nobreak}

\def\address#1{\par\kern 5pt\titlestyle{\twelvepoint\it #1}}
\def\andaddress{\par\kern 5pt \centerline{\sl and} \address}
\def\SLAC{\address{Stanford Linear Accelerator Center\break
      Stanford University, Stanford, California, 94305}}
\def\abstract{\vskip\frontpageskip\centerline{\fourteenrm ABSTRACT}
              \vskip\headskip }
\def\submit#1{\par\nobreak\vfil\nobreak\medskip
   \centerline{Submitted to \sl #1}}
\def\doeack{\foot{Work supported by the Department of Energy,
      contract $\caps DE-AC03-76SF00515$.}}
%
%
%
\def\ie{\hbox{\it i.e.}}

\def\\{\relax\ifmmode\backslash\else$\backslash$\fi}
\def\globaleqnumbers{\relax\ifnum\the\equanumber<0%
\else\global\equanumber=-1\fi}
\def\nextline{\unskip\nobreak\hskip\parfillskip\break}

\def\journal#1&#2(#3){\unskip, \sl #1~\bf #2 \rm (19#3) }
\def\cropen#1{\crcr\noalign{\vskip #1}}
\def\crr{\cropen{10pt}}
\def\topspace{\hrule height 0pt depth 0pt \vskip}

\let\int=\intop         
\def\prop{\mathrel{{\mathchoice{\pr@p\scriptstyle}{\pr@p\scriptstyle}{
                \pr@p\scriptscriptstyle}{\pr@p\scriptscriptstyle} }}}
\def\pr@p#1{\setbox0=\hbox{$\cal #1 \char'103$}
   \hbox{$\cal #1 \char'117$\kern-.4\wd0\box0}}
\def\lsim{\mathrel{\mathpalette\@versim<}}
\def\gsim{\mathrel{\mathpalette\@versim>}}
\def\@versim#1#2{\lower0.2ex\vbox{\baselineskip\z@skip\lineskip\z@skip
  \lineskiplimit\z@\ialign{$\m@th#1\hfil##\hfil$\crcr#2\crcr\sim\crcr}}}
%
%
%
\let\sec@nt=\sec
\def\sec{\relax\ifmmode\let\n@xt=\sec@nt\else\let\n@xt\section\fi\n@xt}
\def\obsolete#1{\message{Macro \string #1 is obsolete.}}
\def\firstsec#1{\obsolete\firstsec \section{#1}}
\def\firstsubsec#1{\obsolete\firstsubsec \subsection{#1}}
\def\thispage#1{\obsolete\thispage \global\pagenumber=#1\frontpagefalse}
\def\thischapter#1{\obsolete\thischapter \global\chapternumber=#1}
\def\nextequation#1{\obsolete\nextequation \global\equanumber=#1
   \ifnum\the\equanumber>0 \global\advance\equanumber by 1 \fi}
\def\BOXITEM{\afterassigment\B@XITEM\setbox0=}
\def\B@XITEM{\par\hangindent\wd0 \noindent\box0 }
%

%
\newdimen\xshift      \xshift=0in
\newdimen\xpos        \xpos=0.5in
\newdimen\yshift      \yshift=0in
\newdimen\ypos        \ypos=0.75in

\def\translate#1#2#3{
   \vbox to 0pt{\offinterlineskip
      \kern-#2\hbox to 0pt{\kern#1{#3}\hss}\vss } }
\def\insertimPRESS#1#2#3{%
    \advance\xshift by -\xpos
    \advance\yshift by -\ypos
      \hbox{%
      \translate{\xshift}{\yshift}{\special{mergeug(#3)}}%
     \blankbox{#1}{#2} }}
\def\blankbox#1#2{\vrule width 0pt depth 0pt height #2
                   \vrule height 0pt depth 0pt width #1}

\def\binsertimPRESS#1{%
    \advance\xshift by -\xpos
    \advance\yshift by -\ypos
      \hbox{%
      \translate{\xshift}{\yshift}{\special{mergeug(#1)}}%
       }}
\def\blankbox#1#2{\vrule width 0pt depth 0pt height #2
                   \vrule height 0pt depth 0pt width #1}

%
\newbox\figbox
\newdimen\zero  \zero=0pt
\newdimen\figmove
\newdimen\figwidth
\newdimen\figheight
\newdimen\figrefheight
\newdimen\textwidth
\newtoks\figtoks
\newcount\figcounta
\newcount\figlines
\def\figreset{\global\figmove=\baselineskip \global\figcounta=0
\global\figlines=1 \global\figtoks={ } }
%
%
%

\def\picture#1by#2:#3{\global\setbox\figbox=\vbox{\vskip #1
\hbox{\vbox{\hsize=#2 \noindent #3}}}
\global\setbox\figbox=\vbox{\kern 5pt
\hbox{\kern 5pt \box\figbox \kern 5pt }\kern 7.5pt}
\global\figwidth=1\wd\figbox
\global\figheight=1\ht\figbox
\global\figrefheight=\figheight
\global\textwidth=\hsize
\global\advance\textwidth by - \figwidth }

\def\figtoksappend{\edef\temp##1{\global\figtoks=%
{\the\figtoks ##1}}\temp}
\def\figparmsa#1{\loop \global\advance\figcounta by 1%
\ifnum \figcounta < #1 \figtoksappend{0pt \the\hsize}%
\global\advance\figlines by 1 \repeat }
\newdimen\figstep
\def\figst@p{\global\figstep = \baselineskip}
\def\figparmsb{\loop \ifdim\figrefheight > 0pt%
\figtoksappend{ \the\figwidth \the\textwidth}%
\global\advance\figrefheight by -\figstep%
\global\advance\figlines by 1%
\repeat }
%
%
%
%
%
\def\figtext#1:#2{\figreset \figst@p%
\figparmsa{#1}%
\figparmsb%
\multiply\figmove by #1%
\global\setbox\figbox=\vbox to 0pt{\vskip\figmove\hbox{\box\figbox}\vss}%
\parshape=\the\figlines\the\figtoks\the\zero\the\hsize%
\noindent\rlap{\box\figbox} #2}
%
%
\catcode`@=12 
\message{ by V.K.}
\everyjob{\input myphyx }

%% file: defs.tex
\hsize=160mm\vsize=240mm
\hoffset=0mm\voffset=0mm
\def\ifm#1{\relax\ifmmode#1\else$#1$\fi}
\def\eps{\ifm{\epsilon}} \def\epm{\ifm{e^+e^-}}
\def\rep{\ifm{\Re(\eps'/\eps)}}    
\def\DAF{DA$\Phi$NE}  
\def\gam{\ifm{\gamma}} \def\to{\ifm{\rightarrow}}
\def\pip{\ifm{\pi^+}} 
\def\po{\ifm{\pi^0}} 
\def\pic{\ifm{\pi^+\pi^-}} \def\pio{\ifm{\pi^0\pi^0}} 
\def\ks{\ifm{K_S}} \def\kl{\ifm{K_L}} \def\kls{\ifm{K_{L,\,S}}} 
 \def\ko{\ifm{K^0}}
\def\K{\ifm{K}} 
\def\Kb{\ifm{\rlap{\kern.3em\raise1.9ex\hbox to.6em{\hrulefill}} K}}
\def\ab{\ifm{\sim}}  \def\x{\ifm{\times}}

\def\sta#1{\ifm{|\,#1\,\rangle}} 
\def\amp#1,#2,{\ifm{\langle#1|#2\rangle}}
\def\kob{\ifm{\Kb\vphantom{K}^0}}
\def\f{\ifm{\phi}}   
  
\def\up#1{$^{#1}$}  \def\dn#1{$_{#1}$}
\def\etal{{\it et al.}}

\def\deg{\ifm{^\circ}}

\def\figure#1]#2]#3]{\par\vbox to #1cm{\vfill\centerline{
\tenpoint{\bf Fig. \chapterlabel#2.} #3} }}

\def\pt#1,#2,{#1\x10\up{#2}}

%% file: subsec.tex
\newcount\subsecno           \subsecno=0
\def\section#1{\par \ifnum\the\lastpenalty=30000\else
   \penalty-200\vskip\sectionskip  \spacecheck\sectionminspace\fi
   \wlog{\string\section\ \chapterlabel \the\sectionnumber}
   \global\advance\sectionnumber by 1  \noindent
   \subsecno=0  
   {\caps\enspace\chapterlabel \sectionlabel #1}\par
   \nobreak\vskip\headskip \penalty 30000 }
\def\subsec#1{\par 
   \global\advance\subsecno by 1  \noindent 
   {\it\enspace\chapterlabel 
    \number\sectionnumber.\number\subsecno \quad #1}\par 
   \nobreak\smallskip } 

%% file: etaf.tex
{\newdimen\unit
\def\legen#1 #2 #3 {\rlap{\kern#1\unit
                \raise#2\unit\hbox{#3}}}

\hbox{\unit=1cm
\rlap{\cl{\epsfig{file=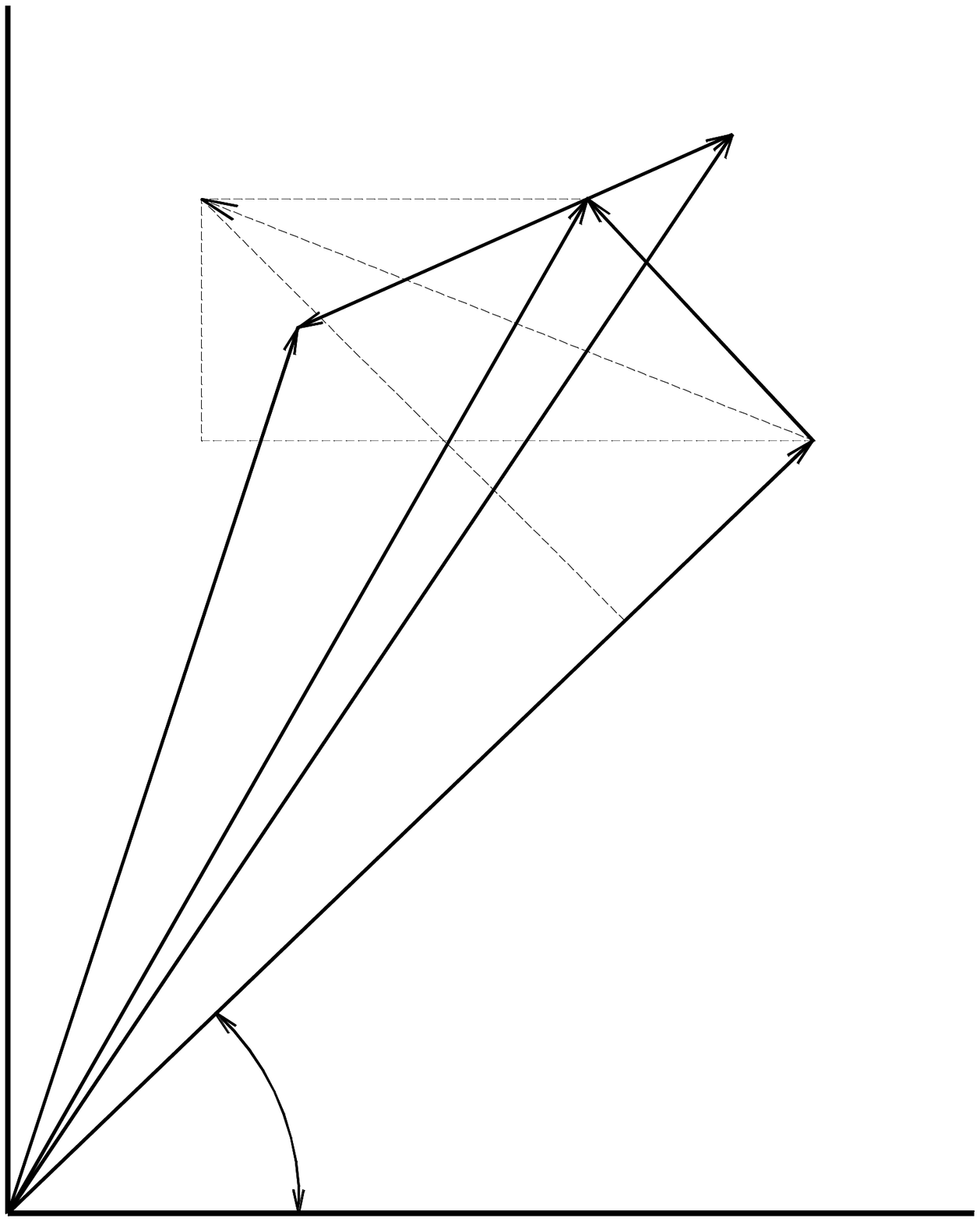,width=8cm}}}
\legen 10 .4 {$\Re$ axis}
\legen 3.7 9.2 {$\Im$ axis}
\legen 5.9 1 {$\phi_{\rm SW}$}
\legen 7.8 5.6 {{\tenpoint$\displaystyle{M-\overline M\over2\sqrt2\Delta m}$}}
\legen 9.3 7.2 {{\tenpoint$-\Delta+\displaystyle{\Re B_0\over A_0}$}}
\legen 8.5 6.9 {{\tenpoint$-\Delta$}}
\legen 5.8 6.4 {{\tenpoint$\Re\Delta$}}
\legen 4.4 7.1 {{\tenpoint$\Im\Delta$}}
\legen 6.2 8.7 {{\tenpoint$\displaystyle{\Re B_0\over A_0}$}}
\legen 6.4 7.8 {$-2\epsilon'$}
\legen 8.6 8.5 {$\epsilon'$}
\legen 6.1 5 {$\epsilon$}
\legen 4.2 4.2 {$\eta_{00}$}
\legen 6.6 4.2 {$\eta_{+-}$}
\legen 7.9 3.7 {$\epsilon_M$}
}
}

%% file: etarho.tex
{\newdimen\unit
\def\legen#1 #2 #3 {\rlap{\kern#1\unit
                \raise#2\unit\hbox{#3}}}

\elvpoint

 \setbox0=\vbox{\baselineskip=11pt
      \ialign{#\hfil\cr
               \kl\to\po\epm\cr
               \kl\to\po$\nu\bar\nu$\cr
               $\eps'/\eps$\cr}}
 \setbox1=\vbox to\ht0{\vfil\hbox{$\bigg\{$}\vfil}

\hbox{\unit=1cm
\rlap{\cl{\epsfig{file=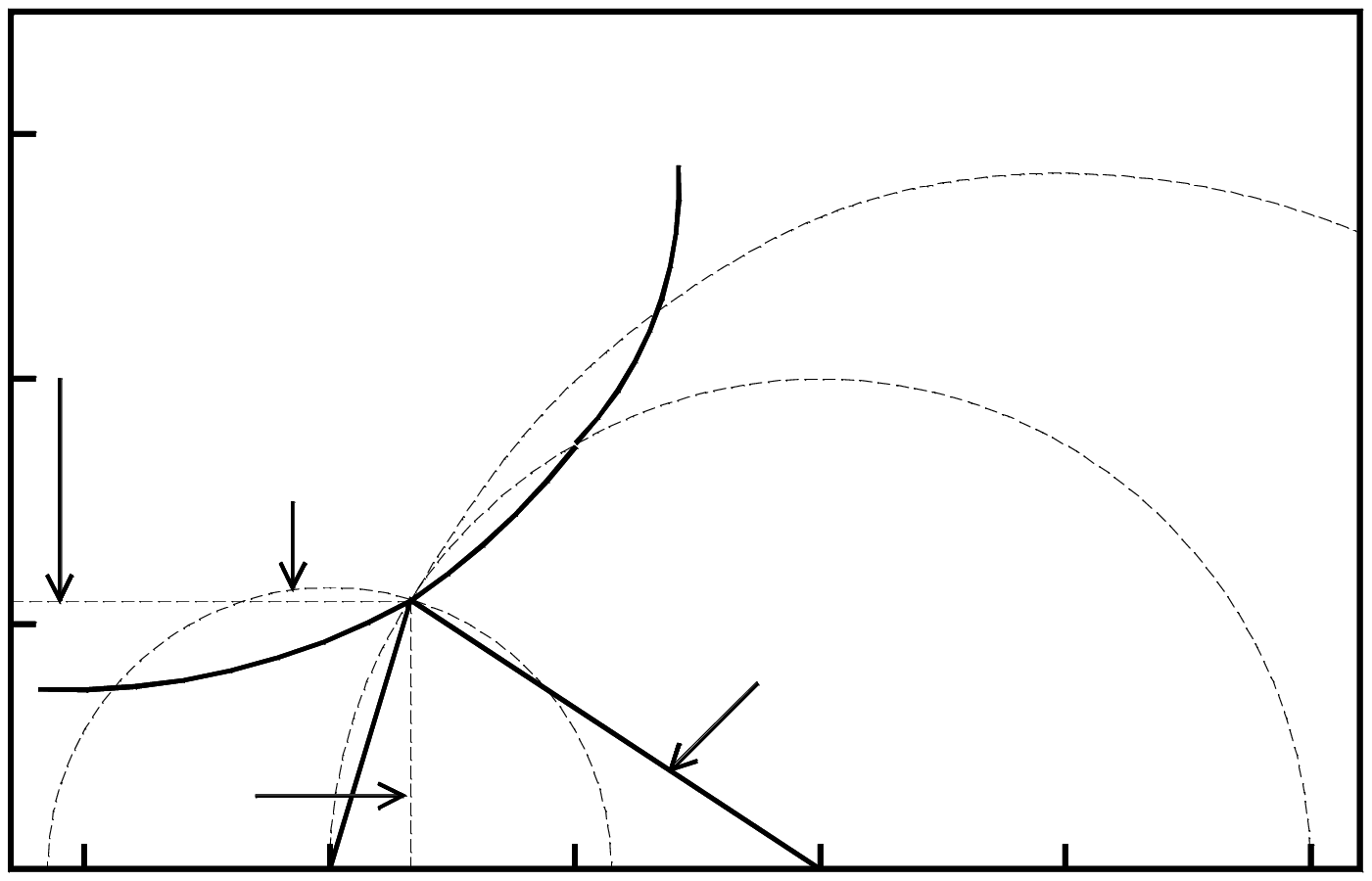,width=12cm}}}
\legen 2.1 3.75 {\hbox{\box1\box0}}
\legen 2.4 1.1 {\kl\to$\mu\bar\mu]$\dn{\rm SD}}
\legen 1.9 1.9 {\eps}
\legen 7.4 5.9 {\eps}
\legen 10.1 4.1 {$B^0$$\leftrightarrow$$\bar B^0$, etc.}
\legen 4.1 3.4 {$\Big|\displaystyle{V_{ub}\over V_{cb}}\Big|$}
\legen 9.6 6.3 {$K^+\to\pi^+\nu\bar\nu$}
\legen 6.6 2 {``Unitarity triangle''}
\legen 1.95 -.3 {-0.5}
\legen 4.25 -.3 {0}
\legen 6.15 -.3 {0.5}
\legen 8.2 -.3 {1.0}
\legen 10.25 -.3 {1.5}
\legen 11.3 -.3 {$\rho$}
\legen 12.25 -.3 {2.5}
\legen .9 2.1 {0.5}
\legen .9 4.15 {1.0}
\legen .9 6.2 {1.5}
\legen 1.1 5.2 {$\eta$}
}

}